\documentclass[12pt, a4paper,oneside, reqno]{amsart} 

\usepackage{amssymb,amsmath}
\usepackage{graphicx,psfrag}

\newcommand{\parrow}{\xrightarrow{\resizebox{!}{3.5pt}{$\circ$}}}

\newcommand{\mrm}{\mathrm}

\newcommand{\vege}{\hfill$\lhd$}
\newcommand{\Ob}{\mrm{Ob}}

\newcommand{\Ph}{\mrm{Ph}}
\newcommand{\F}{\mrm{F}}
\newcommand{\B}{\mrm{B}}
\newcommand{\W}{\mrm{W}}

\newcommand{\IOb}{\mrm{IOb}}

\newcommand{\lam}{\lambda}

\usepackage{amsthm} 
\usepackage{color} 
\definecolor{thmcolor}{rgb}{0,0,.4} 
\definecolor{remarkcolor}{rgb}{0,.2,0} 
\definecolor{proofcolor}{rgb}{.4,0,0} 
\definecolor{quecolor}{rgb}{.2,.2,0} 
\definecolor{axcolor}{rgb}{.23,0,.23}

\renewcommand{\qedsymbol}{\textcolor{proofcolor}{$\blacksquare$}} 
 
\newcommand{\x}[1]{\textcolor{axcolor}{\ensuremath{\mathsf{#1}}}} 
 
\theoremstyle{definition} \newtheorem{thm}{\textcolor{thmcolor}{Theorem}}[section] 
\theoremstyle{definition} \newtheorem{col}[thm]{\textcolor{thmcolor}{Corollary}} 
\theoremstyle{definition} \newtheorem{lem}[thm]{\textcolor{thmcolor}{Lemma}} 
\theoremstyle{definition} \newtheorem{prop}[thm]{\textcolor{thmcolor}{Proposition}} 
 
\theoremstyle{remark} \newtheorem{conv}[thm]{{\sc \textcolor{remarkcolor}{Convention}}} 
\theoremstyle{definition} \newtheorem{que}[thm]{{\textcolor{quecolor}{Question}}} 
\theoremstyle{definition} \newtheorem{rem}[thm]{{\textcolor{remarkcolor}{Remark}}}

\begin{document}

\begin{abstract}
We study the foundation of space-time theory in the framework
of first-order logic (FOL). Since the foundation of mathematics has
been successfully carried through (via set theory) in FOL, it is not
entirely impossible to do the same for space-time theory (or
relativity). First we recall a simple and streamlined
FOL-axiomatization \x{Specrel} of special relativity from the
literature. \x{Specrel} is complete  with respect to 
questions about inertial
motion. Then we ask ourselves whether we can prove the usual
relativistic properties of accelerated motion (e.g., clocks in
acceleration) in \x{Specrel}. As it turns out, this is practically
equivalent to asking whether \x{Specrel} is strong enough to
``handle'' (or treat) accelerated observers. We show that there is a
mathematical principle called induction (\x{IND}) coming from real
analysis which needs to be added to \x{Specrel} in order to handle
situations involving relativistic acceleration. We present an
extended version \x{AccRel} of \x{Specrel} which is strong enough to
handle accelerated motion, in particular, accelerated observers.
Among others, we show that the Twin Paradox becomes provable
in \x{AccRel}, but it is not provable without \x{IND}.

\bigskip

\noindent
Key words: twin paradox, relativity theory, accelerated observers,
first-order logic, axiomatization, foundation of relativity theory
\end{abstract}

\title{TWIN PARADOX AND THE LOGICAL FOUNDATION OF RELATIVITY THEORY}

\author{Judit X.\ Madar\'asz \and Istv\'an N\'emeti \and Gergely Sz\'ekely}

\thanks{Alfr\'ed R\'enyi Institute of Mathematics of the Hungarian Academy of
Sciences,  POB 127
H-1364 Budapest, Hungary.
E-mail addresses: madarasz@renyi.hu, nemeti@renyi.hu,
turms@renyi.hu.}

\date{} 
\maketitle

%%%%%%%%%%%%%%%%%%%%%%%
\section{INTRODUCTION}%
%%%%%%%%%%%%%%%%%%%%%%%
%\medskip

The idea of elaborating the foundation of space-time (or
foundation of relativity) in a spirit analogous with the rather
successful foundation of mathematics (FOM) was initiated by several
authors including, e.g., David Hilbert~\cite{Hi02} or leading contemporary
logician Harvey Friedman~\cite{FriFOM1, FriFOM2}. 
Foundation of mathematics has been carried through strictly
within the framework of first-order logic (FOL), for certain reasons. The
same reasons motivate the effort of keeping the foundation of
space-time also inside FOL. One of the reasons is that staying
inside FOL helps us to avoid tacit assumptions, another reason is
that FOL has a complete inference system while higher-order logic
cannot have one by G\"odel's incompleteness theorem, see e.g., 
V{\"a}{\"a}n{\"a}nen~\cite[p.505]{V01} or \cite[Appendix]{pezsgo}. For more
motivation for staying inside FOL (as opposed to higher-order logic),
cf.\ e.g., Ax \cite{Ax}, Pambuccian \cite{Pam},
\cite[Appendix 1: ``Why exactly FOL'']{pezsgo}, \cite{AMNsamp}, but
the reasons in V\"a\"an\"anen \cite{V01}, Ferreir\'os \cite{FeBSL},
or Wole\'nski \cite{Wol} also apply.

Following the above motivation, we begin at the beginning, namely
first we recall a streamlined FOL axiomatization \x{Specrel} of
special relativity theory, from the literature. \x{Specrel} is
complete with respect to (w.r.t.) questions about inertial motion. Then we ask
ourselves whether we can prove the usual relativistic properties of
accelerated motion (e.g., clocks in acceleration) in \x{Specrel}. As
it turns out, this is practically equivalent to asking whether
\x{Specrel} is strong enough to ``handle'' (or treat) accelerated
observers. We show that there is a mathematical principle called
induction (\x{IND}) coming from real analysis which needs to be
added to \x{Specrel} in order to handle situations involving
relativistic acceleration. We present an extended version \x{AccRel}
of \x{Specrel} which is strong enough to handle accelerated clocks,
in particular, accelerated observers.

We show that the so-called Twin Paradox becomes provable in
\x{AccRel}. It also becomes possible to introduce Einstein's
equivalence principle for treating gravity as acceleration and
proving the gravitational time dilation, 
i.e.\ that gravity ``causes time to run
slow''.

What we are doing here is not unrelated to Field's ``Science
without numbers'' programme and to ``reverse mathematics'' in the
sense of Harvey Friedman and Steven Simpson. Namely, we
systematically ask ourselves which mathematical principles or
assumptions (like, e.g., \x{IND}) are really needed for proving certain
observational predictions of relativity. (It was this striving for
parsimony in axioms or assumptions which we alluded to when we
mentioned, way above, that \x{Specrel} was ``streamlined''.)

The interplay between logic and relativity theory goes back to
around 1920 and has been playing a non-negligible role in works of
researchers like Reichenbach, Carnap, Suppes, Ax, Szekeres,
Malament, Walker, and of many other contemporaries.%
\footnote{ In passing we mention that Etesi-N\'emeti~\cite{EN},
Hogarth~\cite{Hdecid} represent further kinds of
\emph{connection} between \emph{logic and relativity} not
discussed here.}

In Section~\ref{ax-s} we recall the FOL axiomatization \x{Specrel}
complete w.r.t.\ questions concerning inertial motion. There we also
introduce an extension \x{AccRel} of \x{Specrel} (still inside FOL)
capable for handling accelerated clocks and also accelerated
observers. In Section~\ref{main-s} we formalize the Twin Paradox in the
language of FOL. We formulate Theorems~\ref{thmTwp}, \ref{thmEq} stating that the
Twin Paradox is provable from \x{AccRel} and the same for related
questions for accelerated clocks. Theorems \ref{thmNoIND}, \ref{thmMO} state that
\x{Specrel} is not sufficient for this, more concretely that the
induction axiom \x{IND} in \x{AccRel} is needed. In Sections~\ref{an-s},
\ref{proofss} we
prove these theorems.

Motivation for the research direction reported here is nicely
summarized in Ax~\cite{Ax}, Suppes~\cite{Sup68}; cf.\ also the
introduction of \cite{pezsgo}. Harvey Friedman's \cite{FriFOM1, FriFOM2}
present a rather convincing general perspective (and
motivation) for the kind of work reported here.

%\medskip
%%%%%%%%%%%%%%%%%%%%%%%%%%%%%%%%%%%%%%%%%%%%%%%%%
\section{AXIOMATIZING SPECIAL RELATIVITY IN FOL}%
%%%%%%%%%%%%%%%%%%%%%%%%%%%%%%%%%%%%%%%%%%%%%%%%%
%\medskip

\label{ax-s}

In this paper we deal with the kinematics of relativity only,
i.e.\ we deal with motion of {\em bodies} (or {\em test-particles}).
The motivation for our choice of vocabulary (for special relativity)
is summarized as follows. We will represent motion as changing
spatial location in time. To do so, we will have reference-frames
for coordinatizing events and, for simplicity, we will associate
reference-frames with special bodies which we will call {\em
observers}. We visualize an observer-as-a-body as ``sitting'' in the
origin of the space part of its reference-frame, or equivalently, 
``living'' on the
time-axis of the reference-frame. We will distinguish {\em inertial}
observers from non-inertial (i.e.\ accelerated) ones. There will be
another special kind of bodies which we will call {\em photons}. For
coordinatizing events we will use an arbitrary {\em ordered field}
in place of the field of the real numbers. Thus the elements of this
field will be the ``{\em quantities}'' which we will use for marking
time and space. Allowing arbitrary ordered fields in place of
the reals increases flexibility of our theory and minimizes the
amount of our mathematical presuppositions.  Cf.\ e.g., Ax \cite{Ax}
for further motivation in this direction. Similar remarks apply to
our flexibility oriented decisions below, e.g., keeping the number
$d$  of space-time dimensions a variable. Using coordinate systems
(or reference-frames) instead of a single observer independent
space-time structure is only a matter of didactical convenience and
visualization, furthermore it also helps us in weeding out unnecessary
axioms from our theories. Motivated by the above, we now turn to
fixing the FOL language of our axiom systems.

The first occurrences of concepts used in this work are set by
boldface letters to make it easier to find them. Throughout this
work, if-and-only-if is abbreviated to {\bf iff}.

Let us fix a natural number $d\ge 2$ for the dimension of the space-time that  we are going to axiomatize.
Our first-order language contains the following non-logical symbols:
\begin{itemize}
\item unary relation symbols $\B$ (for {\bf Bodies}), $\Ob$ (for {\bf Observers}),
$\IOb$ (for {\bf Inertial Observers}), $\Ph$ (for {\bf Photons}) and $\F$ (for {\bf quantities} which are going to be elements of a Field),
\item binary function symbols  $+$, $\cdot$ and a binary relation symbol $\le $ (for the field operations and the ordering on
$\F$), and
\item a $2+d$-ary relation symbol $\W$ (for {\bf World-view relation}).
\end{itemize}

The bodies will play the role of the
``main characters''  of our space-time models
and they will be ``observed'' (coordinatized using the
quantities) by the observers. This observation will be coded by the
world-view relation $\W$. Our bodies and observers are basically the same as
the ``test particles'' and the ``reference-frames'', respectively, in some of the literature.

We read $\B(x), \Ob(x), \IOb(x), \Ph(x), \F(x)$ as ``$x$ is a body'',
``$x$ is an observer'', ``$x$ is an inertial observer'', ``$x$ is a photon'', ``$x$ is a
field-element''.
We use the world-view  relation $\W$ to talk about coordinatization, by reading
$\W(x,y,z_1,\ldots, z_d)$ as ``observer $x$ observes (or sees) 
body $y$ at coordinate point
$\langle z_1,\ldots,z_d\rangle$''.
This kind of observation has no connection with seeing via photons, it simply means
coordinatization.

$\B(x), \Ob(x), \IOb(x), \Ph(x), \F(x), \W(x,y,z_1,\ldots, z_d), x=y, x\leq y$
are the so-called {atomic formulas} of our first-order
language, where $x,y,z_1,\dots,z_d$ can be arbitrary variables or terms built
up
from variables by using the field-operations ``$+$'' and ``$\cdot$''. The
{\bf formulas} of  our first-order language are built up from these
atomic formulas by using the logical connectives {\em not}
($\lnot$), {\em and} ($\land$), {\em or} ($\lor$), {\em implies}
($\Longrightarrow$), {\em if-and-only-if} ($\Longrightarrow$) and the
quantifiers {\em exists} $x$ ($\exists x$) and {\em for all $x$} ($\forall x$)
for every variable $x$.

Usually we  use the variables $m,k,h$ to denote observers, $b$ to
denote bodies,  $ph$ to denote photons and
$p_1,\dots,q_1,\dots$ to denote quantities (i.e.\ field-elements).
We write $p$ and $q$ in place of $p_1,\dots,p_d$ and
$q_1,\dots,q_d$, e.g., we  write $\W(m,b,p)$ in place of
$\W(m,b,p_1,\dots,p_d)$,  and we  write $\forall p$ in place of
$\forall p_1,\dots,p_d$ etc.

The {\bf models} of this language are  of the form
\begin{equation}
\mathfrak{M} = \langle U; \B, \Ob, \IOb, \Ph, \F,+,\cdot,\leq,\W\rangle,
\end{equation}
where $U$ is a nonempty set,
$\B,\Ob,\IOb,\Ph,\F$ are unary relations on $U$, etc. A unary
 relation on
$U$ is just a subset of $U$.
Thus we   use $\B,\Ob$ etc.\ as sets as
well, e.g., we  write $m\in \Ob$ in place of $\Ob(m)$.

\smallskip

Having fixed our language, we now turn to formulating an axiom
system for special relativity in this language. We will make special
efforts to keep all our axioms inside the above specified
first-order logic language of $\mathfrak{M}$.

Throughout this work, $i$, $j$ and $n$ denote positive integers.
$\F^n:=\F\times\ldots\times \F$
($n$-times) is the set of all $n$-tuples of elements of $\F$.
If  $a\in \F^n$, then we  assume that $a=\langle a_1,\ldots,a_n\rangle$, i.e.\
$a_i\in\F$ denotes the $i$-th component of the $n$-tuple $a$.

The following  axiom is always  assumed and is part of every axiom system we 
propose.

\begin{description}
\item[\x{AxFrame}] $\Ob\cup \Ph\subseteq \B$, $\IOb\subseteq \Ob$,
$U=\B\cup \F$, $\W\subseteq \Ob \times \B\times \F^d$, $+$ and
$\cdot$ are binary operations on $\F$, $\le$ is a binary relation on
$\F$ and $\left< \F; +,\cdot, \le \right>$ is an {\bf Euclidean
ordered field}, i.e.\ a linearly ordered field in which positive
elements have square roots.%
\footnote{ For example, the ordered fields of the real numbers,  the
real algebraic numbers, and the hyper-real numbers are Euclidean
but the ordered field  of the
rational numbers is not Euclidean and the field
of the complex numbers cannot be ordered. For the definition of
(linearly) ordered field, cf.\ e.g., Rudin~\cite{Rudin} or
Chang-Keisler~\cite{Chang-Keisler}.}

\end{description}

In pure first-order logic, the above axiom would look like $\forall
x\enskip\big[\big(\Ob(x)\vee \Ph(x)\big)\Longrightarrow \B(x)\big]$ 
etc. In the present section
we will not write out the purely first-order logic
translations of our axioms since they will be 
straightforward to obtain. The first-order logic
translations of our next three axioms \x{AxSelf^-}, \x{AxPh}, \x{AxEv}
can be found in the Appendix.

Let $\mathfrak{M}$ be a model in which \x{AxFrame} is true.  Let
$\mathfrak{F}:=\left< \F; +, \cdot, \le\right>$ denote the {\bf
ordered field reduct} of $\mathfrak{M}$. Here we list the
definitions and notation 
that we are going to use in formulating our axioms. Let
$0,1,-,/,\sqrt{\phantom{i}}$ be the usual field operations which are
definable from ``$+$'' and ``$\cdot$''.
We use the vector-space structure of $\F^n$,  i.e.\ if $p,q\in \F^n$
and $\lambda\in \F$, then $p+q,-p, \lambda p\in \F^n$; and
$o:=\langle 0,\ldots,0\rangle$ denotes the {\bf origin}. The {\bf
Euclidean-length} of $a\in \F^n$ is defined as { $|a|
:=\sqrt{\resizebox{!}{8pt}{$a_1^2+\ldots+a_n^2$}}$}. The set of
positive elements of $\F$ is denoted by $\F^+:=\{x\in \F:x>0\}$.
Let $p,q\in \F^d$.
We use the notation   $p_s:=\langle p_2,\ldots, p_d\rangle$ for  the
{\bf space component} of $p$ and $p_t:=p_1$ for the {\bf time component} of $p$.
We define the {\bf line} through  $p$ and $q$ as
$pq:=\{ q+\lam(p-q): \lam \in \F\}$.
The set of {\bf lines} is defined as
$\text{\it Lines}:=\{pq : p\neq q \;\land\; p,q \in \F^d\}$.
The {\bf slope} of $p$ is defined as
$\text{\it slope}(p):=
{\left| p_s  \right|}/{\left| p_t \right|}$
if $p_t\neq 0$ and is undefined otherwise; furthermore
$\text{\it slope}(pq):=\text{\it slope}(p-q)$
if $p_t\neq q_t$ and is undefined otherwise.
$\F^d$ is called the {\bf coordinate system} and its elements are referred to as {\bf coordinate points}.
The {\bf event}  (the set of bodies) observed by observer $m$  at coordinate point $p$ is:
\begin{equation}
ev_m(p):=\{b\in \B : \W(m,b,p)\}.
\end{equation}
The mapping $p\mapsto ev_m(p)$ is called the {\bf world-view} (function) of $m$.
The {\bf coordinate domain}  of  observer $m$ is the set of coordinate points where $m$
observes something:
\begin{equation}
Cd(m):=\{p \in \F^d: ev_m(p)\neq \emptyset  \}.
\end{equation}
The {\bf life-line} (or {\bf trace}) of body $b$ as seen by observer $m$ is defined as the set of coordinate points where $b$ was observed by $m$:
\begin{equation}
tr_m(b):=\{p\in \F^d : \W(m,b,p)\}=\{p\in \F^d : b \in ev_m(p)\}.
\end{equation}
The life-line $tr_m(m)$ of  observer $m$ as seen by
himself is called the {\bf self-line} of $m$. The {\bf time-axis} is
defined as $\bar{t}:=\{\langle x,0,\dots,0\rangle:x\in \F\}$.

\begin{figure}[h!btp]
\small
\begin{center}
\psfrag{mm}[bl][bl]{$tr_m(m)$}
\psfrag{mk}[bl][bl]{$tr_m(k)$}
\psfrag{mb}[t][t]{$tr_m(b)$}
\psfrag{mph}[tl][tl]{$tr_m(ph)$}
\psfrag{kk}[br][br]{$tr_k(k)$}
\psfrag{km}[bl][bl]{$tr_k(m)$}
\psfrag{kb}[t][t]{$tr_k(b)$}
\psfrag{kph}[tl][tl]{$tr_k(ph)$}
\psfrag{p}[r][r]{$p$}
\psfrag{q}[l][l]{$q$}
\psfrag{t}[lb][lb]{$\bar{t}$}
\psfrag{o}[t][t]{$o$}
\psfrag{fkm}[t][t]{$ev_k(p)=ev_m(q) \text{, i.e.\ } \langle p,q\rangle
\in f^k_m$}
\psfrag*{text1}[cb][cb]{world-view of $k$}
\psfrag*{text2}[cb][cb]{world-view of $m$}
\includegraphics[keepaspectratio, width=0.8\textwidth]{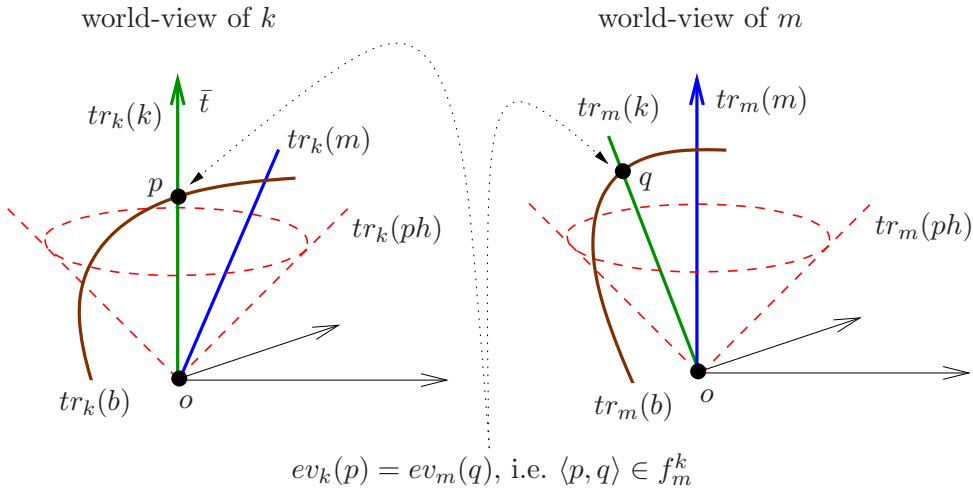}
\caption{for the basic definitions mainly for $f^k_m$. }
\end{center}
\end{figure}

Now we are ready to build our space-time theories by formulating our axioms.
We formulate each axiom on two levels.
First we give an intuitive formulation, then we give a
precise formalization using our notation.

The following natural axiom goes back to Galileo Galilei and even to the Norman-French
Oresme of around 1350, cf.\ e.g., \cite[p.23, \S 5]{AMNsamp}. It simply states that each
observer thinks that he rests in the origin of the space part of his coordinate system.
\begin{description}
\item[\x{AxSelf^-}] The self-line of any observer is the  time-axis
restricted to his coordinate domain:
\begin{equation}
\forall m \in \Ob\quad tr_m(m)=\bar{t}\cap Cd(m).
\end{equation}
\end{description}
A FOL-formula expressing \x{AxSelf^-} can be found in the Appendix. 

The next axiom is about the constancy of the speed of the photons, cf.\ e.g., 
\cite[\S 2.6]{d'Inverno}.
For convenience, we choose  $1$ for their speed.
\begin{description}
\item[\x{AxPh}]  For every inertial observer, the lines of slope 1 are exactly the traces of the photons:
\begin{equation}
\forall m\in \IOb\quad\{tr_{m}(ph):ph\in \Ph\}=\{l\in \text{\it
Lines}:\text{\it slope}(l)=1\}.
\end{equation}
\end{description}
A FOL-formula expressing \x{AxPh} can be found in the Appendix.

We will also assume the following axiom:
\begin{description}
\item[\x{AxEv}] All inertial observers observe the same 
events:
\begin{equation}
\forall m,k\in \IOb\enskip \forall p\in \F^d\;\exists q\in \F^d\quad ev_{m}(p)=ev_{k}(q).
\end{equation}
\end{description}
A FOL-formula expressing \x{AxEv} can be found in the Appendix.

\begin{equation}
\x{Specrel_0}:= \{ \x{AxSelf^-}, \x{AxPh}, \x{AxEv}, \x{AxFrame}\}.
\end{equation}

\medskip

Since, in some sense, \x{AxFrame} is only an ``auxiliary'' (or
book-keeping) axiom about the ``mathematical frame'' of our
reasoning, the heart of \x{Specrel_0} consists of three very natural
axioms, \x{AxSelf^-}, \x{AxPh}, \x{AxEv}. These are really
intuitively convincing, natural and simple assumptions. From these
three axioms one can already prove the most characteristic
predictions of special relativity theory. What the average layperson
usually knows about relativity is that ``moving clocks slow down'',
``moving spaceships shrink'', and ``moving pairs of clocks get out
of synchronism''. We call these the {\bf paradigmatic effects} of
special relativity. All these can be proven from the above three
axioms, in some form, cf.\ Theorem~\ref{spec-thm}. E.g., one can prove
that ``if $m,k$ are any two observers not at rest relative to each
other, then one of $m,k$ will ``see'' or ``think'' that the clock of
the other runs slow''.
However, \x{Specrel_0} does not imply yet the inertial approximation
of the so-called Twin Paradox.%
\footnote{ This inertial approximation of the twin paradox is formulated 
as \x{AxTp^{in}} at the end of Section~\ref{main-s} below
Theorem~\ref{thmMO}.}
 In order to prove the inertial
approximation of the Twin Paradox also, and to prove all the
paradigmatic effects in their strongest form, it is enough to add
one more axiom \x{AxSym} to \x{Specrel_0}. This is what we are going
to do now.

We will find that studying the relationships between the
world-views is more illuminating than studying the world-views in
themselves. Therefore  the following
definition is fundamental. The {\bf world-view transformation}
between the world-views of observers $k$ and $m$  is the set of pairs of
coordinate points $\langle p,q\rangle$ 
such that $m$ and $k$ observe the same nonempty 
event in $p$ and $q$, respectively:
\begin{equation}
f^k_m:=\{\langle p,q\rangle \in \F^d \times \F^d:ev_k(p)=ev_m(q)\neq\emptyset\},
\end{equation}
cf.\ Figure~1. We note that although the world-view
transformations are only binary relations,  axiom
\x{AxPh} turns them into functions, cf.\ (iii) of
Proposition~\ref{prop0} way below.

\begin{conv}\label{fmkconv}
Whenever we write ``$f^k_m(p) $'', we mean that there is a  unique
$q\in \F^d$ such that $\langle p,q\rangle \in f^k_m$, and $f^k_m(p)$
denotes this unique $q$. I.e.\ if we talk about the value
$f^k_m(p)$ of $f^k_m$ at $p$, we tacitly postulate that it exists
and is unique (by the present convention). 
\end{conv}

The following axiom is an instance (or special case) of
the Principle of Special Relativity,
according to which the ``laws of nature'' are the same for
all inertial observers, in particular,
there is no experiment which would decide whether you are in
absolute motion, cf.\ e.g., Einstein \cite{Einstein} or \cite[\S
2.5]{d'Inverno} or \cite[\S 2.8]{Mphd}. To explain the following
formula, let $p,q\in\F^d$. Then $p_t-q_t$ is the time passed between
the events $ev_m(p)$ and $ev_m(q)$ as seen by $m$ and
$f^m_k(p)_t-f^m_k(q)_t$ is the time passed between the same two
events as seen by $k$. Hence $|( f^m_k(p)_t-
f^m_k(q)_t)\slash(p_t-q_t)|$ is the rate with which $k$'s clock runs
slow as seen by $m$. The same explanation applies when  $m$  and  $k$  
are interchanged.

\begin{description}
\item[\x{AxSym}]  Any two inertial 
observers see each other's clocks go wrong in the same way:
\begin{equation}
\forall m,k\in \IOb\enskip \forall  p,q\in
\bar{t}\quad\big|f^k_m(p)_t - f^k_m(q)_t\big|=\big| f^m_k(p)_t-
f^m_k(q)_t\big|.
\end{equation}
\end{description}

All the axioms so far talked about inertial observers, and
they in fact form an axiom system complete w.r.t.\ the inertial
observers, cf.\ Theorem \ref{spec-thm} below.

\begin{equation}
\x{Specrel}:= \{ \x{AxSelf^-}, \x{AxPh}, \x{AxEv}, \x{AxSym}, \x{AxFrame}\}.
\end{equation}

\medskip

Let $p,q\in \F^d$. Then
\begin{equation}
\mu(p):=\left\{
\begin{array}{rl}
\sqrt{\big|p_t^2-|p_s|^2 \big|}  & \text{ if }  p_t^2-|p_s|^2\ge0 , \\
-\sqrt{\big|p_t^2-|p_s|^2 \big|} & \text{ otherwise }
\end{array}
\right.
\end{equation}
is the (signed) {\bf Minkowski-length} of $p$ and the {\bf
Minkowski-distance} between  $p$ and $q$ is defined as follows:
\begin{equation}
\mu(p,q):=\mu(p-q).
\end{equation}
A motivation for the ``otherwise'' part of the
definition of  $\mu(p)$  is the following.  $\mu(p)$  codes two kinds of
information, (i) the length of  $p$  and (ii) whether  $p$  is time-like 
(i.e.\ $|p_t|>|p_s|$) or space-like. Since the length is always non-negative, 
we can use the sign of $\mu(p)$  to code  (ii).

Let $f:\F^d\rightarrow \F^d$ be a function.
$f$ is said to be a {\bf Poincar\'e-transformation} if
$f$ is a bijection and it preserves the
Minkowski-distance, i.e.\ $\mu\big(f(p),f(q)\big)=\mu(p,q)$ for all
$p,q\in \F^d$.
$f$ is called a {\bf dilation} if 
there is a
positive $\delta\in\F$ such that $f(p)=\delta p$ for all
$p\in\F^d$ and $f$ is called a {\bf field-automorphism-induced} mapping if
there is an
automorphism $\pi$ of the field $\langle\F,+,\cdot\rangle$ such that
$f(p)=\langle \pi p_1,\dots, \pi p_d\rangle$ for all $p\in\F^d$.
The following is proved in \cite[2.9.4, 2.9.5]{pezsgo} and in
\cite[2.9.4--2.9.7]{Mphd}.

Let 
$\Sigma$ be a set of formulas and  
$\mathfrak{M}$ be a model.
 $\mathfrak{M} \models  \Sigma $   denotes that all formulas in 
$\Sigma$ are true
in  model  $\mathfrak{M}$.
In this case we say that  $\mathfrak{M}$ is a {\bf model of} $\Sigma$.

\begin{thm} \label{spec-thm}\label{thmPoi} Let $d>2$, let $\mathfrak{M}$ 
be a model of
our language and let $m,k$ be inertial observers in $\mathfrak{M}$. Then
$f^m_k$ is a Poincar\'e-transformation whenever
$\mathfrak{M}\models\x{Specrel}$. More generally,  $f^m_k$ is a
Poincar\'e-transformation composed with a dilation and a
field-automorphism-induced mapping  whenever
$\mathfrak{M}\models\x{Specrel_0}$.
 \end{thm}

\begin{rem}
\label{rem-specthm}
Assume $d>2$. Theorem~\ref{spec-thm} is best possible in the
sense that, e.g., for every Poincar\'e-transformation $f$ over an
arbitrary Euclidean ordered field there are a model
$\mathfrak{M}\models\x{Specrel}$ and inertial observers $m,k$ in
$\mathfrak{M}$ such
that the world-view transformation $f^m_k$ between $m$'s and $k$'s
world-views in $\mathfrak{M}$ is $f$, see \cite[2.9.4(iii),
2.9.5(iii)]{pezsgo}. Similarly for the second statement in
Theorem~\ref{spec-thm}. Hence, Theorem \ref{spec-thm} can be refined to a
pair of completeness theorems, cf.\ \cite[3.6.13, p.271]{pezsgo}.
Roughly, for every Euclidean ordered field, its
Poincar\'e-transformations (can be expanded to) form a model of
\x{Specrel}. Similarly for the other case.
\vege
\end{rem}

It follows from Theorem \ref{spec-thm} that the paradigmatic
effects of relativity hold in \x{Specrel} in their strongest form,
e.g., if $m$ and $k$ are observers not at rest w.r.t.\ each other,
then both will ``think'' that the clock of the other runs slow.
\x{Specrel} also implies the ``inertial approximation'' of the Twin
Paradox,
see e.g., \cite[2.8.18]{pezsgo}, and \cite{mythes}. 
It is
necessary to add \x{AxSym} to \x{Specrel_0} in order to be able to
prove the inertial approximation of the Twin Paradox by
Theorem \ref{spec-thm}, cf.\ e.g., \cite{mythes}.
\bigskip

We now begin to formulate axioms about non-inertial
observers. The non-inertial observers are called {\bf accelerated}
observers. To connect the world-views of the accelerated and the
inertial observers, we are going to formulate the statement that at
each moment of his life, each accelerated observer sees the nearby
world for a short while as an inertial observer does. To
formalize this, first we introduce the
relation of being a co-moving observer. 
 The (open) {\bf ball} with center
$c\in \F^n$ and radius $\varepsilon \in \F^+$ is:
\begin{equation}
B_\varepsilon(c):=\{a\in \F^n : \left| a-c \right| < \varepsilon\}.
\end{equation}
 $m$ is a {\bf co-moving observer} of $k$ at $q\in \F^d$, in symbols
 $m \succ_q k$, if  $q\in Cd(k)$ and the following holds:
\begin{equation}
\forall \varepsilon \in \F^+ \;  \exists \delta \in \F^+ \enskip
\forall p \in B_{\delta}(q)\cap Cd(k)\enskip \big|p-f^k_m(p)\big|
\leq\varepsilon|p-q|.
\end{equation}

Behind the definition of the co-moving observers is the following
intuitive image: as we zoom into smaller and smaller neighborhoods
of the given coordinate point, the world-views of  the two observers
are more and more similar. Notice that $f^k_m(q)=q$ if $m \succ_q k$.

The following axiom gives the promised connection between the world-views of
the inertial and the accelerated observers:

\begin{description}
\item[\x{AxAcc}] At any point on the self-line of any observer, 
there is a co-moving inertial observer:
\begin{equation}
\forall k \in \Ob \enskip \forall q \in tr_k(k) \; \exists m \in \IOb\quad m\succ_q k.
\end{equation}
\end{description}

Let \x{AccRel_0} be the collection of the axioms introduced so far:
\begin{equation}
\x{AccRel_0}:=\{ \x{AxSelf^-}, \x{AxPh}, \x{AxEv}, \x{AxSym}, \x{AxAcc},
\x{AxFrame}\}.
\end{equation}

\medskip

Let $\mathfrak{R}$ denote the ordered field of the real
numbers.
Accelerated clocks behave as expected in models of \x{AccRel_0}
when the ordered field reduct of the model is 
$\mathfrak{R}$ (cf.\ Theorems~\ref{thmTwp}, 
\ref{thmEq}, and in more detail Prop.\ref{propTr}, Rem.\ref{Trrem}); 
but not otherwise (cf.\ Theorems~\ref{thmNoIND}, \ref{thmMO}).
Thus to prove properties of accelerated clocks (and
observers), we need more properties of the field reducts than their
being Euclidean ordered fields. As it turns out, adding all
FOL-formulas valid in the field of the reals does not suffice (cf.\
Corollary~\ref{corNoIND}). 
The additional property of $\mathfrak{R}$ we need is that
in $\mathfrak{R}$ every bounded non-empty set has a {\bf supremum},
i.e.\ a least upper bound. This is a second-order logic property  (because
it concerns all subsets of $\mathfrak{R}$) which
we cannot use in a FOL axiom system. Instead, we will use a kind of
``induction'' axiom schema. It will state that every non-empty,
bounded subset of the ordered field reduct which can be defined by a
FOL-formula using possibly the extra part of the model, e.g., using the
world-view relation, has a supremum. We now begin to formulate our FOL
induction axiom schema.%
\footnote{ This way of imitating a second-order formula by a
FOL-formula schema comes from the methodology of approximating
second-order theories by FOL ones, examples are Tarski's replacement
of Hilbert's second-order geometry axiom by a FOL schema or Peano's
FOL induction schema replacing second-order logic induction.}

To abbreviate formulas of FOL we often omit parentheses according to the
following convention. Quantifiers bind as long as they can, and $\land$
binds stronger than $\Longrightarrow$. E.g., $\forall x\
\varphi\land\psi\Longrightarrow\exists y\ \delta\land\eta$ means $\forall
x\big((\varphi\land\psi)\Longrightarrow\exists y(\delta\land\eta)\big)$. 
Instead of
curly brackets we sometimes write square brackets in formulas, e.g., we
may write $\forall x\ (\varphi\land\psi \Longrightarrow[\exists
y\delta]\land\eta)$.

 If $\varphi$ is a formula and $x$ is
a variable, then we say that $x$ is a {\bf free variable} \label{free
variable} of
$\varphi$ if{}f $x$ does not occur under the scope of either
$\exists x$ or $\forall x$. 

Let $\varphi$ be a formula; and let $x,y_1,\ldots,y_n$ be all the
free variables of $\varphi$. Let
$\mathfrak{M}=\langle U;\B,\ldots\rangle$ be a model. Whether
$\varphi$ is true or false in $\mathfrak{M}$ depends on how we
associate elements of $U$ to these free variables. When we
associate $d,a_1,\ldots,a_n\in U$ to $x,y_1,\ldots,y_n$,
respectively, then $\varphi(d,a_1,\ldots,a_n)$ denotes this
truth-value, thus $\varphi(d,a_1,\ldots,a_n)$ is either true or
false in $\mathfrak{M}$. 
For example, if $\varphi$ is $x\leq
y_1+\ldots+y_n$, then $\varphi(0,1,\ldots,1)$ is true in
$\mathfrak{R}$ while $\varphi(1,0,\ldots,0)$ is false in
$\mathfrak{R}$. 
$\varphi$ is said to be {\bf true} in $\mathfrak{M}$ 
if $\varphi$ is true in
$\mathfrak{M}$ no matter how we associate elements to the free 
variables.
We say that 
a subset $H$ of $\F$ is  {\bf definable
by} $\varphi$ iff there are $a_1,\ldots,a_n\in U$ such
that $H=\{d\in \F\: :\: \varphi(d,a_1,\ldots,a_n)\text{ is true in
}\mathfrak{M}\}$.

\goodbreak

\begin{description}
\item[\x{AxSup_\varphi}] Every subset
of $\F$ definable by  $\varphi$
has  a supremum if it is non-empty and {\bf bounded}:
\begin{equation}
\begin{split}
\forall y_1,\ldots,y_n  \quad[\exists x \in \F \quad \varphi] \;\land\;
[\exists b\in \F \quad (\forall x\in \F\quad \varphi \Longrightarrow x \le
b)]\;\\ \Longrightarrow\; 
\big(\exists s \in \F \enskip \forall b \in \F\quad (\forall x\in \F\quad
\varphi \Longrightarrow x \le b)\iff s\le b\big).
\end{split}
\end{equation}
\end{description}
We say that a
subset of $\F$ is {\bf definable} if{}f it is definable by a
FOL-formula. Our axiom scheme \x{IND} below says that every non-empty
bounded and definable subset of $\F$ has a supremum.

\begin{equation}
\x{IND}:=\{\x{AxSup_\varphi}:    \varphi \mbox{ is a FOL-formula of 
our language }  \}.
\end{equation}

Notice that \x{IND} is true in any model whose ordered field reduct is 
$\mathfrak{R}$.
Let us add \x{IND} to \x{AccRel_0} and call it \x{AccRel}:

\begin{equation}
\x{AccRel}:=\x{AccRel_0}\cup\x{IND}.
\end{equation}

\x{AccRel} is a countable set of FOL-formulas. We note that
there are non-trivial models of \x{AccRel}, cf.\ e.g., Remark~\ref{Trrem}
way below. Furthermore, we note
that the construction in 
Misner-Thorne-Wheeler~\cite[Chapter 6 entitled ``The local coordinate
system of an accelerated observer'', especially pp.\ 172-173 and Chapter
13.6 entitled ``The proper reference frame of an accelerated observer'' on
pp.\ 327-332]{MTW} 
 can be used for constructing models of \x{AccRel}.
Models of \x{AccRel} are discussed to
some detail in \cite{mythes}. Theorems~\ref{thmTwp} and \ref{thmEq}
(and also Prop.\ref{propTr}, Rem.\ref{Trrem})
below show that \x{AccRel} already implies properties of
accelerated clocks, e.g., it implies the Twin Paradox.

\x{IND} implies all the FOL-formulas true in $\mathfrak{R}$, 
but \x{IND} is stronger. Let \x{IND^-}
denote the set of elements of \x{IND} that talk only about the
ordered field reduct, i.e.\ let
\begin{equation}
\begin{split}
\x{IND^-} := \{ & \x{AxSup_\varphi} : 
        \varphi 
\mbox{ is a FOL-formula in the language } \\ 
&\mbox{of the ordered field reduct of our models}\}.
\end{split}
\end{equation}

Now, \x{IND^-} together with the axioms of  linearly ordered fields is a 
complete axiomatization of the FOL-theory  of $\mathfrak{R}$, i.e.\ all 
FOL-formulas valid in $\mathfrak{R}$
can be derived from them.%
\footnote{ This follows from a theorem of Tarski, cf.\
Hodges~\cite[p.68 (b)]{Ho97} or Tarski~\cite{Ta51},
by Theorem~\ref{thmBoltzano} herein or \cite[Proposition A.0.1]{mythes}.} 
However, \x{IND} is
stronger than \x{IND^-}, since  
$\x{AccRel_0}\cup\x{IND^-} \not\models \x{Tp}$ by Corollary~\ref{corNoIND} 
below,
while  $\x{AccRel_0}\cup\x{IND}\models \x{Tp}$ by Theorem~\ref{thmTwp}. 
The strength of \x{IND}  comes from
the fact that the formulas in \x{IND} can ``talk'' 
about more ``things'' than just
those in the language of $\mathfrak{R}$ 
(namely they can talk about the world-view
relation, too). For understanding how  \x{IND}  works, 
it is important to notice
that \x{IND} does not speak about the field $\mathfrak{F}$  
itself, but instead, it speaks
about connections between  $\mathfrak{F}$
and the rest of the model $\mathfrak{M}=\langle\ldots, \F,\ldots,\W\rangle$.

Why do we call \x{IND} a kind of induction schema? 
The reason is the following.
\x{IND} implies that if a formula  $\varphi$  
becomes false sometime after $0$  while
being true at $0$, then there is a ``first'' time-point where, 
so to speak,  $\varphi$ becomes false. This time-point is the supremum of 
the time-points until which  $\varphi$
remained true after $0$. Now, $\varphi$ may or may not be 
false at this supremum,
but it is false arbitrarily ``close'' to it afterwards. If such a ``point of
change'' for the truth of  $\varphi$  cannot exist, \x{IND} implies that  
$\varphi$ has to be
true always after $0$ if it is true at $0$. (Without \x{IND}, 
this may not be true.)

%\medskip
%%%%%%%%%%%%%%
\section{MAIN RESULTS: ACCELERATED CLOCKS AND THE 
TWIN PARADOX IN OUR FOL AXIOMATIC SETTING}%
%%%%%%%%%%%%%%
%\medskip

\label{main-s}

Twin Paradox (TP) concerns two twin siblings whom we shall
call Ann and Ian. (``A'' and ``I'' stand for accelerated and for
inertial, respectively). Ann travels in a spaceship to some distant
star while Ian remains at home. TP states that when Ann returns home
she will be \emph{younger} than her \emph{twin brother} Ian. We now
formulate TP in our FOL language.

The
{\bf segment} between $p\in \F^d$ and $q\in \F^d$ is defined as:
\begin{equation}
[p q]:=\{\lam p+(1-\lam)q:\lambda\in \F\;\land\; 0\le\lam\le1\}.
\end{equation}

We say that observer $k$ is in {\bf twin-paradox relation} with
observer $m$ iff whenever $k$ leaves $m$ between two
meetings, $k$ measures less time between the two meetings than $m$:
\begin{equation}
\begin{split}
\forall p,q \in tr_k(k) \enskip \forall p',q' \in tr_m(m) &  
\\ \langle
p,p'\rangle ,\langle q,q'\rangle \in f^k_m \;\land\;[p q]
\subseteq & tr_k(k)\;\land\; [p' q']\not\subseteq tr_m(k)\\
& \Longrightarrow\;  \big|q_t-p_t\big|<\big|q'_t-p'_t\big|,
\end{split}
\end{equation}
cf.\ Figure~2.
In this case we write $\mathsf{Tp}(k<m)$. We note that, if two observers do not leave each other or they meet less than twice, then they are in twin-paradox relation by this definition. Thus two inertial observers are always in this relation.

\begin{figure}[h!btp]
\small
\begin{center}
\psfrag*{TwP}[l][l]{\x{Tp}}
\psfrag*{Eqtime}[l][l]{\x{Ddpe}}
\psfrag*{f}[b][b]{$f^k_m$}
\psfrag*{p}[rt][rt]{$p$}
\psfrag*{q}[rb][rb]{$q$}
\psfrag*{t1}[l][l]{$tr_m(k)$}
\psfrag*{p'}[lt][lt]{$p'$}
\psfrag*{q'}[lb][lb]{$q'$}
\psfrag*{a}[c][c]{$\Longrightarrow\; |q_t-p_t|<|q'_t-p'_t|$}
\psfrag*{b}[c][c]{$\Longrightarrow\; |q_t-p_t|=|q'_t-p'_t|$}
\psfrag*{m}[b][b]{$tr_m(m)$}
\psfrag*{k}[b][b]{$tr_k(k)$}
\psfrag*{t3}[r][r]{$ tr_m(k) \not\supseteq$}
\psfrag*{t2}[r][r]{$ tr_k(k) \supseteq$}
\psfrag*{c}[c][c]{same events}
\includegraphics[keepaspectratio, width=0.8\textwidth]{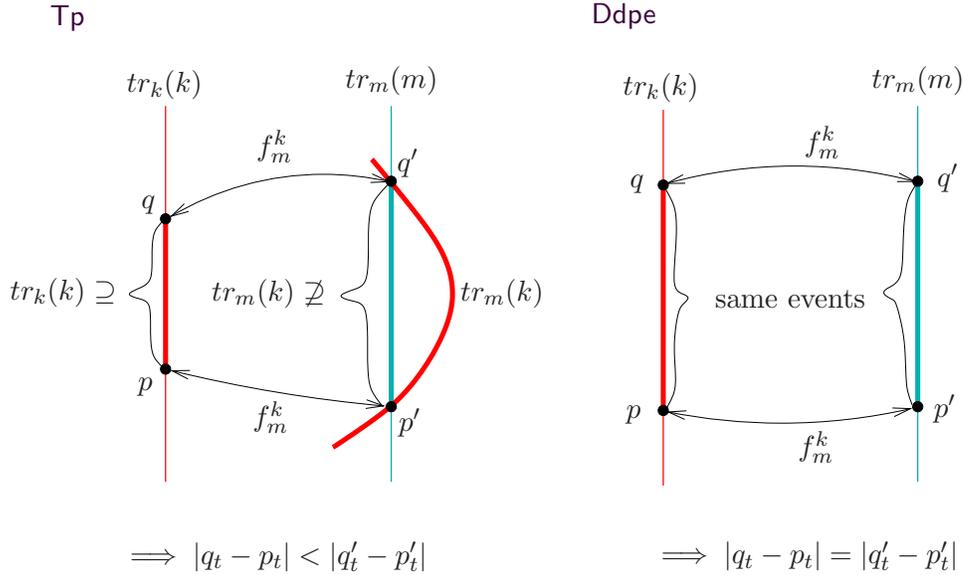}
\caption{for \x{Tp} and \x{Ddpe}. }
\end{center}
\end{figure}

\begin{description}
\item[\x{Tp}] Every observer is in twin-paradox relation with every inertial observer:
\begin{equation}\forall k\in \Ob\enskip\forall m\in \IOb\quad \mathsf{Tp}(k<m).
\end{equation}
\end{description}

Let $\varphi$ be a formula and $\Sigma$ be a set of formulas.
$\Sigma \models \varphi$ denotes that $\varphi$ is true in all  models 
of $\Sigma$.
G\"odel's completeness theorem for FOL implies that whenever
$\Sigma\models\varphi$, there is a (syntactic) derivation of
$\varphi$ from $\Sigma$ via the commonly used derivation rules of
FOL. 
Hence the next theorem states that the formula \x{Tp} formulating the
Twin Paradox is provable
from the axiom system \x{AccRel}.

\begin{thm} \label{thmTwp}
$\x{AccRel} \models \x{Tp}$ if $d>2$. 
\end{thm}
\noindent
The proof of the theorem is in Section~\ref{proofss}.

\bigskip

Now we turn to formulating a phenomenon which we call Duration
Determining Property of Events.
\begin{description}
\item[\x{Ddpe}] If each of
two observers observes the very same (non-empty) events in a segment of their
self-lines,
they measure the same time  between the end points of these two segments:
\begin{equation}
\begin{split}
\forall k,m\in \Ob\enskip\forall p,q \in tr_k(k)\enskip \forall p',q'\in
t&r_m(m)\\
\emptyset\not\in\{ev_k(r):r\in[p q]\}=\{ev_m(r'&):  r'\in [p' q']\}\;
\Longrightarrow\;\\ \big|&q_t-p_t\big|=\big|q'_t-p'_t\big|,
\end{split}
\end{equation}
see the right hand side of Figure~2.
\end{description}

The next theorem states that \x{Ddpe} also can be proved from
our FOL axiom system \x{AccRel}.

\begin{thm} \label{thmEq}
$\x{AccRel} \models \x{Ddpe}$ if $d>2$.
\end{thm}
\noindent
The proof of the theorem is in Section~\ref{proofss}.

\begin{rem} The assumption $d>2$ cannot be omitted from
Theorem~\ref{thmTwp}. However, Theorems~\ref{thmTwp} and \ref{thmEq} remain true if we
omit the assumption $d>2$ and assume  auxiliary axioms \x{AxIOb} and \x{AxLine}
below, i.e.\
\begin{equation}
\x{AccRel}\cup\{\x{AxIOb},\x{AxLine}\}\models \x{Tp}\;\land\;\x{Ddpe}
\end{equation}
holds for $d=2$, too.
A proof for the latter statement can be obtained from the proofs 
of Theorems~\ref{thmTwp} and
\ref{thmEq} by \cite[items 4.3.1, 4.2.4, 4.2.5]{mythes}
and \cite[Theorem 1.4(ii)]{AMNsamp}.

\begin{description}
\item[\x{AxIOb}] In every inertial observer's coordinate system, every line of slope less than 1 is
the life-line of
an inertial observer:
\begin{equation}
\forall m\in \IOb\quad \{tr_m(k): k\in \IOb \} \supseteq \{l \in \text{\it
Lines}: \text{\it slope}(l)< 1 \}. 
\end{equation}

\item[\x{AxLine}] Traces of  inertial observers are lines as observed by 
inertial
observers:
 \begin{equation}
\forall m,k \in \IOb\quad tr_m(k)\in \text{\it Lines}.
\end{equation}
\end{description}
\vege
\end{rem}

\begin{que}
Can the assumption $d>2$ be omitted from Theorem~\ref{thmEq}, i.e.\
does $\x{AccRel}\models\x{Ddpe}$ hold for $d=2$?
\end{que}
 
The following theorem says that Theorems~\ref{thmTwp} and \ref{thmEq} do
not remain true if we omit the axiom scheme \x{IND} from \x{AccRel}.
If a formula $\varphi$ is not true in 
a model $\mathfrak{M}$, we write
$\mathfrak{M} \not\models \varphi$.

\begin{thm}\label{thmNoIND}
For every Euclidean ordered field $\mathfrak{F}$ not isomorphic to 
$\mathfrak{R}$, there
is a model
$\mathfrak{M}$ of $\x{AccRel_0}$ such that
$\mathfrak{M}\not\models\x{Tp}$, $\mathfrak{M}\not\models\x{Ddpe}$ 
and the ordered field
reduct of $\mathfrak{M}$ is $\mathfrak{F}$.
\end{thm}
\noindent The proof of the theorem is in Section~\ref{proofss}. 
\bigskip

By Theorems~\ref{thmTwp} and \ref{thmEq}, 
\x{IND} is not true in the model $\mathfrak{M}$ mentioned in 
Theorem~\ref{thmNoIND}.
This theorem has
 strong consequences, it implies that to prove the Twin Paradox,
it does not suffice to add all the FOL-formulas valid in $\mathfrak{R}$ 
(to \x{AccRel_0}). Let  $Th(\mathfrak{R})$  denote the set of all FOL-formulas valid in
$\mathfrak{R}$. 

\begin{col}  
\label{corNoIND}
$Th(\mathfrak{R})\cup\x{AccRel_0}\not\models \x{Tp}$  
and  $Th(\mathfrak{R})\cup\x{AccRel_0}\not\models \x{Ddpe}$.
\end{col}
\noindent The proof of the corollary is in Section~\ref{proofss}. 
\bigskip

An ordered field is called {\bf non-Archimedean} if it has an element $a$ such
that, for every
positive integer $n$, $-1<\underbrace{a+\ldots+a}_n<1$.
We call these elements {\bf infinitesimally small}.

The following theorem says that, for countable or non-Archimedean
Euclidean ordered fields, there are quite
sophisticated models of \x{AccRel_0} in which \x{Tp} and \x{Ddpe}
are false.

\begin{thm}\label{thmMO}
For every Euclidean ordered field $\mathfrak{F}$ which is countable or
non-Archimedean, there is a  model $\mathfrak{M}$  of $\x{AccRel_0}$ such
that $\mathfrak{M}\not\models\x{Tp}$, $\mathfrak{M}\not\models\x{Ddpe}$, 
the ordered field
reduct of $\mathfrak{M}$ is $\mathfrak{F}$ and (i)--(iv) below
also hold in $\mathfrak{M}$.
\begin{itemize}
\item[(i)] Every observer uses the whole coordinate system for
coordinate-domain:
\begin{equation}
\forall m \in \Ob\quad Cd(m)=\F^d.
\end{equation}
\item[(ii)] At any point in $\F^d$, there is a co-moving inertial observer of any
observer:
\begin{equation}
\forall k \in \Ob \enskip \forall q \in \F^d\; \exists m \in \IOb\quad m\succ_q k.
\end{equation}
\item[(iii)] All observers observe the same set of events:
\begin{equation}
\forall m,k\in \Ob\enskip \forall p\in \F^d\;\exists q\in \F^d\quad ev_{m}(p)=ev_{k}(q).
\end{equation}
\item[(iv)] Every observer observes every event only once:
\begin{equation}
\forall m\in \Ob\enskip \forall p,q\in \F^d\quad ev_m(p)=ev_m(q)\;
\Longrightarrow\; p=q. 
\end{equation}
\vege
\end{itemize}
\end{thm}
\noindent

The proof of the theorem is in Section~\ref{proofss}.

\bigskip

\begin{figure}[h!btp]
\small
\begin{center}
\psfrag{m}[l][l]{$tr_m(m)$}
\psfrag{k2}[rt][rt]{$tr_{k_2}(k_2)$}
\psfrag{k1}[rb][rb]{$tr_{k_1}(k_1)$}
\psfrag{t1}[rt][rt]{$tr_m(k_1)$}
\psfrag{t2}[rb][rb]{$tr_m(k_2)$}
\psfrag{p'}[l][l]{$p'$}
\psfrag{q'}[l][l]{$q'$}
\psfrag{r'}[l][l]{$r'$}
\psfrag{f1}[lb][lb]{$f^{k_2}_m$}
\psfrag{f2}[lb][lb]{$f^{k_1}_m$}
\psfrag{f1t}[t][t]{$f^{k_2}_m$}
\psfrag{f2b}[b][b]{$f^{k_1}_m$}
\psfrag{p}[r][r]{$p$}
\psfrag{q}[r][r]{$q$}
\psfrag{r}[r][r]{$r$}
\includegraphics[keepaspectratio, width=0.5\textwidth]{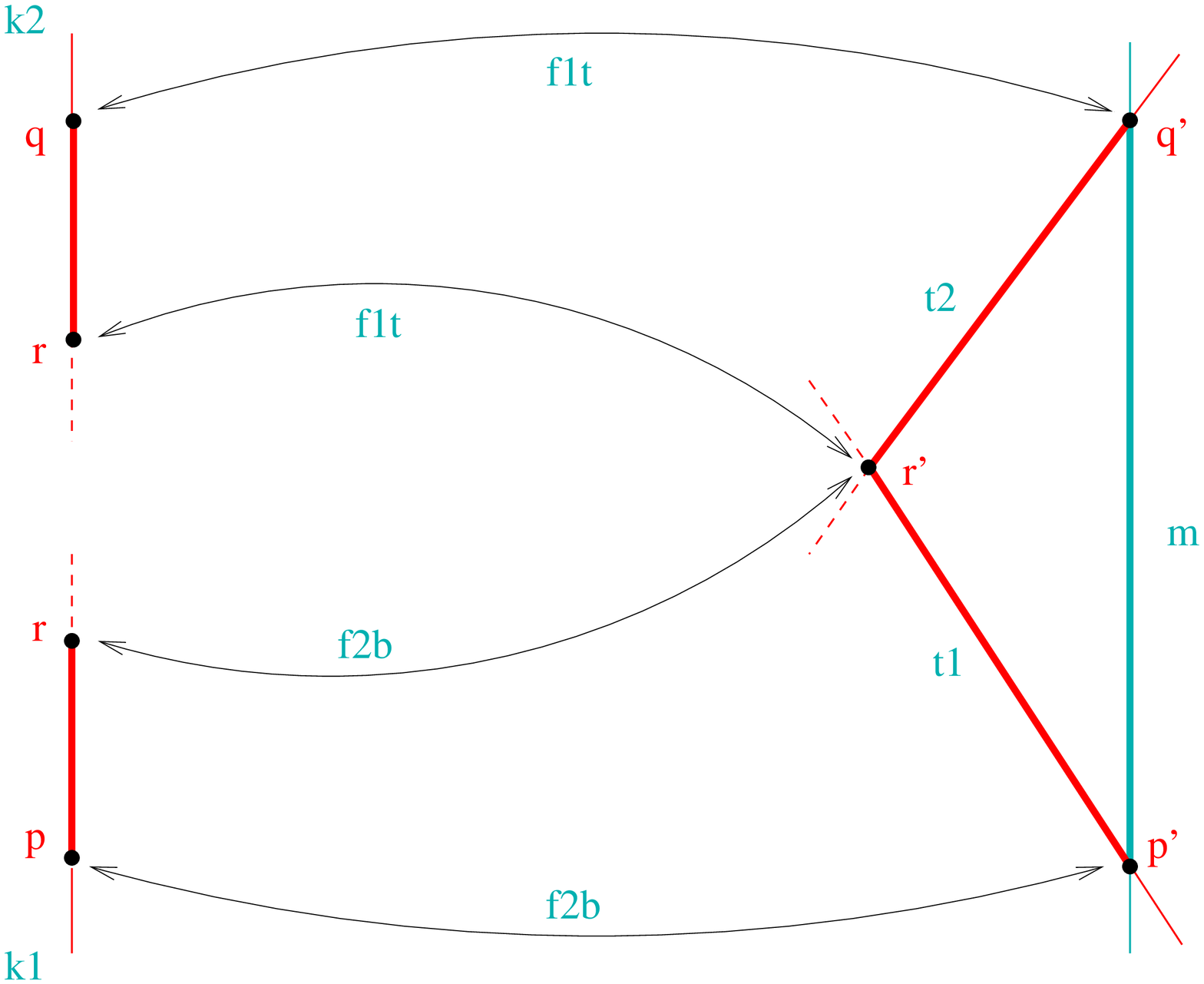}
\caption{for \x{AxTp^{in}}. }
\end{center}
\end{figure}

Finally we formulate a question. To this end we introduce the
inertial version of the twin paradox and some auxiliary axioms. In
the inertial version of the twin paradox, we use the common trick of
the literature to talk about the twin paradox without talking about
accelerated observers. We replace the accelerated twin with two
inertial ones, a leaving and an approaching one.

We say that observers $k_1$ and  $k_2$ are  in {\bf inertial twin-paradox relation} with  observer $m$ if
the following holds:
\begin{equation}
\begin{split}
\forall p,q,r \in tr_{k_1}(k_1)\cap tr_{k_2}(k_2) \enskip \forall p',q' \in
tr&_m(m)\enskip
\forall r'\in
\F^d \quad \\
\langle p,p'\rangle ,\langle r,r'\rangle \in f^{k_1}_m\;\land\;
\langle r,r'\rangle ,\langle q,q'\rangle \in f&^{k_2}_m\;\land\;
p'_t<r'_t<q'_t\;\land\;
 r'\not\in [p' q'] \\
\Longrightarrow& \; |q'_t-p'_t| > |q_t-r_t|+|r_t-p_t\big|,
\end{split}
\end{equation}
cf.\ Figure~3.
In this case we write $\mathsf{Tp}({k_1k_2}<m)$.

\begin{description}
\label{Axintwp}
\item[\x{AxTp^{in}}] Every three inertial observers are in inertial twin-paradox relation:
\begin{equation}
\forall m,k_1,k_2\in \IOb \quad \mathsf{Tp}({k_1k_2}<m).
\end{equation}
\end{description}

\begin{description}
\item[\x{AxTrn}] To every inertial observer $m$ and coordinate point $p$ there is an  inertial
observer $k$ such that the world-view transformation between $m$ and $k$ is the translation  by
vector $p$:
\begin{equation}
\forall m\in \IOb\enskip  \forall p\in \F^d\; \exists k\in \IOb\enskip \forall q\in
\F^d \quad f^m_k(q)=q+p.
\end{equation}
\end{description}

\begin{description}
\item[\x{AxLt}] The world-view transformation between  inertial
observers $m$ and $k$ is a linear transformation if $f^m_k(o)=o$\ :
\begin{equation}
\begin{split}
\forall m,k\in \IOb\enskip\forall p,q\in \F^d\enskip\forall\lambda\in
\F\quad
f^m_k(o)=o\; &\Longrightarrow\;\\ f^m_k(p+q)=f^m_k(p)+f^m_k(q)\;\land\; f^m_k(\lambda
p)=&\lambda f^m_k(p).
\end{split}
\end{equation}
\end{description}

\begin{que}\label{qTwp}
Does Theorem~\ref{thmTwp} remain true if we replace \x{AxSym} in
\x{AccRel} with the inertial version \x{AxTp^{in}} of the twin paradox and
we assume the
auxiliary axioms \x{AxLt}, \x{AxIOb} and 
\x{AxTrn}? Cf.\ Question~\ref{qConv}.
We note that \x{AxTp^{in}} and \x{AxLt} are true in the models
of \x{Specrel} in case  $d>2$, cf.\ \cite[Theorem 1.2]{AMNsamp},
\cite[Theorem 2.8.28]{Mphd} and \cite[\S 3]{mythes}.
\vege
\end{que}

%\medskip
%%%%%%%%%%%%%%
\section{PIECES FROM NON-STANDARD ANALYSIS: SOME TOOLS FROM REAL ANALYSIS 
CAPTURED IN FOL}%
%%%%%%%%%%%
%\medskip

\label{an-s}

In this section we gather the statements (and proofs from
\x{AccRel}) of the facts we will need from analysis. The point is in
formulating these statements in FOL and for an arbitrary ordered
field in place of using the second-order language of the ordered
field $\mathfrak{R}$ of reals.

In the present section \x{AxFrame} is assumed without any further mentioning.

 Let $a,b,c\in \F$. We say that
$b$ is {\bf between} $a$ and $c$ iff $a<b<c$ or $a>b>c$. We use the
following notation: $[a,b]:=\{x\in \F: a\le x \le b\}$ and
$(a,b):=\{x\in \F: a<x<b\}$.
\begin{conv}
Whenever we write $[a,b]$, we assume that $a,b\in \F$ and $a\leq b$.
We also use this convention for $(a,b)$. 
\end{conv}
Let $H\subseteq \F^n$. Then $p\in \F^n$ is said to be an {\bf
accumulation point} of $H$ if for all $\varepsilon\in \F^+$,
$B_\varepsilon(p)\cap H$ has an element different from $p$. $H$ is called {\bf open} if for
all $p\in H$, there is an $\varepsilon\in \F^+$ such that
$B_\varepsilon(p)\subseteq H$. Let $R\subseteq A\times B$ and
$S\subseteq B\times C$ be  binary relations. The {\bf composition}
of $R$ and $S$ is defined as: $ R \circ S :=\{ \langle a,c\rangle
\in A\times C: \exists b\in B \enskip \langle a,b\rangle \in R
\;\land\; \langle b,c\rangle \in S \}$. The {\bf domain} and the
{\bf range} of $R$ are denoted by $Dom(R):=\{a\in A : \exists b\in
B\enskip \langle a,b\rangle \in R \}$ and $Rng(R):= \{b\in B :
\exists a\in A \enskip \langle a,b\rangle\in R \}$, respectively.
$R^{-1}$ denotes the {\bf inverse} of $R$, i.e.\ $R^{-1}:=\{\langle
b,a\rangle \in B\times A: \langle a,b\rangle \in R\}$. We think of a
{\bf function} as a special binary relation. Notice that if $f,g$
are functions, then $f \circ g (x)=g\big(f(x)\big)$ for all $x\in Dom(f\circ
g)$. $f:A\rightarrow
B$ denotes that $f$ is a function from $A$ to $B$, i.e.\ $Dom(f)=A$
and $Rng(f)\subseteq B$. Notation $f:A\parrow B$ denotes that $f$ is
a {\bf partial function} from $A$ to $B$; this means that $f$ is a
function, $Dom(f)\subseteq A$ and $Rng(f)\subseteq B$. Let
$f:\F\parrow \F^n$. We call $f$ {\bf continuous} at  $x$ if $x\in
Dom(f)$, $x$ is an accumulation point of $Dom(f)$ and the usual
formula of continuity holds for $f$ and $x$, i.e.
\begin{equation}
\forall \varepsilon \in \F^+ \; \exists \delta  \in \F^+ \enskip \forall y \in Dom(f)
\quad \left|y - x \right| < \delta
\; \Longrightarrow\; \left|f(y)-f(x) \right| <\varepsilon.
\end{equation}
We call $f$ {\bf differentiable} at $x$ if $x\in Dom(f)$, $x$ is an accumulation point of $Dom(f)$ and there is an $a\in
\F^n$ such that
\begin{equation}
\begin{split}
\forall \varepsilon\in \F^+\; \exists \delta \in \F^+\enskip\forall y\in
Dom(f)&\\
|y-x|<\delta\; \Longrightarrow\; |f(y)-&f(x)-(y-x)a|\le \varepsilon|y-x|.
\end{split}
\end{equation}
This $a$ is unique.
We call this $a$ the {\bf derivate} of $f$ at $x$ and we denote it by $f'(x)$.
$f$ is said to be  continuous (differentiable) on
$H\subseteq \F$ iff $H\subseteq Dom(f)$ and  $f$  is
continuous (differentiable) at every $x\in H$.
We note that the basic properties of the differentiability remain
true since their proofs use only the ordered field properties of $\mathfrak{R}$, cf.\
Propositions~\ref{propDiff}, \ref{propAff} and \ref{propMax} below.

Let $f,g:\F\parrow \F^n$ and $\lambda \in \F$. Then $\lambda f:\F\parrow \F^n$ and
$f+g:\F\parrow \F^n$ are defined as
$\lambda f:=\big\{\langle x,\lambda f(x)\rangle: x \in Dom(f)\big\}$ and
$f+g:=\big\{\langle x,f(x)+g(x)\rangle: x\in Dom(f)\cap Dom(g)\big\}$.
Let $h:\F\parrow \F$.
$h$ is said to be {\bf increasing} on $H$ iff $H\subseteq Dom(h)$ and
for all $x,y\in H$, $h(x)<h(y)$ if $x<y$, and
$h$ is said to be  {\bf decreasing} on $H$ iff $H\subseteq Dom(h)$
and for all $x,y\in H$, $h(x)>h(y)$ if $x<y$.

\goodbreak

\begin{prop}\label{propDiff}
Let $f,g:\F \parrow \F^n$ and $h:\F\parrow \F$.
Then (i)--(v) below hold.
\begin{itemize}
\item[(i)]  If $f$ is differentiable at $x$  then it is also continuous at $x$.
\item[(ii)] Let $\lambda\in \F$. If $f$ is differentiable at $x$, then $\lambda f$ is also differentiable at $x$ and $(\lambda f)'(x)=\lambda f'(x)$.
\item[(iii)]  If $f$ and $g$ are differentiable at $x$ and $x$ is an accumulation point of $Dom(f)\cap
Dom(g)$, then $f+g$ is differentiable at $x$ and $(f+g)'(x)=f'(x)+g'(x)$.
\item[(iv)] If $h$ is differentiable at $x$, $g$ is differentiable at $h(x)$ and $x$ is an accumulation point of
$Dom(h\circ g)$, then
$h\circ g$ is differentiable at $x$ and $(h\circ g)'(x)=h'(x)g'\big(h(x)\big)$.
\item[(v)] If $h$ is increasing (or decreasing) on $(a,b)$, differentiable at $x\in (a,b)$ and
$h'(x)\neq0$, then $h^{-1}$ is differentiable at $h(x)$.
\end{itemize}
\end{prop}

\begin{proof}[\textcolor{proofcolor}{on the proof}]
Since the proofs of the statements are based on  the same calculations and 
ideas  as in 
real analysis, we omit the proof,
cf.\  \cite[Theorems 28.2, 28.3, 28.4 and 29.9]{Ross}.
\end{proof}

Let $i\leq n$. $\pi_i:\F^n\rightarrow \F$ denotes
the $i$-th projection function, i.e.\ $\pi_i:p\mapsto p_i$. Let $f:\F\parrow
\F^n$.
We denote the $i$-th coordinate function of $f$ by $f_i$, i.e.\  $f_i:=f\circ\pi_i$.
We also denote $f_1$ by $f_t$.
A function $A:\F^n\rightarrow \F^j$ is said to be an {\bf affine map}  if it is a linear map composed by a translation.%
\footnote{ I.e.\ $A$ is an affine map if there are $L:\F^n\rightarrow \F^j$ and $a\in \F^j$ such that $A(p)=L(p)+a$,
$L(p+q)=L(p)+L(q)$ and
$L(\lambda p)=\lambda L(p)$ for all $p,q\in \F^n$ and $\lambda \in \F$.}

The following proposition says that the derivate of a function $f$ composed by an affine map $A$ at a point $x$
is the image of the derivate  $f'(x)$ taken by the linear part of $A$.

\begin{prop}\label{propAff}
Let $f:\F \parrow \F^n$ be  differentiable at $x$ and let $A:\F^n\rightarrow
\F^j$ be an affine map.
Then $f\circ A$ is differentiable at $x$ and $(f\circ A)'(x)=A\big(f'(x)\big)-A(o)$.
In particular, $f'(x)=\langle f'_1(x),\dots,f'_n(x)\rangle$, 
i.e.\  $f'_i(x)=f'(x)_i$.
\end{prop}
\begin{proof}[\textcolor{proofcolor}{on the proof}] The statement is straightforward from the definitions.\end{proof}

$f:\F\parrow \F$ is  said to be {\bf locally maximal} at  $x$ iff  $x\in Dom(f)$ and there is
a $\delta \in \F^+$ such that $f(y)\le f(x)$ for all $y\in B_\delta(x)\cap Dom(f)$. The {\bf local minimality} is defined analogously.

\begin{prop}\label{propMax}
If $f:\F\parrow \F$ is differentiable on $(a,b)$ and 
locally maximal or minimal at  $x\in (a,b)$, then its 
derivate is $0$ at $x$, i.e.\ $f'(x)=0$.
\end{prop}

\begin{proof}[\textcolor{proofcolor}{on the proof}] The proof is the same as in real analysis, cf.\
e.g., \cite[Theorem 5.8]{Rudin}.\end{proof}

Let $\mathfrak{M}=\langle U;\ldots\rangle$ be a model.
An $n$-ary relation $R\subseteq \F^n$ is said to be {\bf definable} iff there is a formula $\varphi$ with only
free variables $x_1,\ldots,x_n, y_1,\ldots,y_i$ and there are
$a_1,\ldots,a_i\in U$  such that
\begin{equation}
R=\{\langle p_1,\ldots, p_n\rangle \in \F^n :
\varphi(p_1,\ldots,p_n,a_1,\ldots,a_i) \text{ is true in } \mathfrak{M}\}.
\end{equation}
Recall that  \x{IND} says that  every non-empty, bounded and definable subset of
$\F$ has  a supremum.

\begin{thm}[Bolzano's Theorem]\label{thmBoltzano}
Assume \x{IND}. Let $f: \F \parrow \F$ be definable and  continuous on $[a,b]$. If $c$ is between $f(a)$ and $f(b)$,
then there is an $s\in [a,b]$ such that $f(s)=c$.
\end{thm}

\begin{proof}
Let $c$ be between $f(a)$ and $f(b)$.
We can assume that $f(a)<f(b)$.
Let $H:=\{x\in [a,b] :  f(x) <c\}$. Then $H$ is definable, bounded and non-empty.
Thus, by \x{IND}, it has a supremum, say $s$.
Both  $\{x\in (a,b) : f(x)<c \}$  and $\{x\in (a,b): f(x)>c \}$ are
non-empty open sets since $f$ is continuous on $[a,b]$.
Thus $f(s)$ cannot be less than $c$ since $s$ is an upper bound of $H$ and cannot be greater than $c$ since $s$ is the smallest upper bound. Hence $f(s)=c$ as desired.
\end{proof}

\begin{thm}\label{thmsup}
Assume \x{IND}.  Let $f: \F \parrow \F$ be  definable and  continuous on 
$[a,b]$. Then 
the supremum $s$ of
$\{f(x): x\in [a,b]\}$ exists and there is an
 $y\in [a,b]$ such that $f(y)=s$.
\end{thm}

\begin{proof}
The set  $H:=\{y\in [a,b]: \exists c\in \F\enskip\forall x\in [a,y]\quad f(x)<c\}$ has  
a supremum by \x{IND} since $H$ is definable, non-empty and bounded.
This supremum has to be $b$ and $b\in H$ since $f$ is continuous on $[a,b]$.
Thus $Ran(f):=\{f(x): x\in [a,b]\}$ is bounded. Thus, by \x{IND}, it has a supremum, say $s$, since it is definable and non-empty.
We can assume that $f(a)\neq s$.
Let $A:=\{y\in [a,b]: \exists c \in \F\enskip \forall x\in [a,y]\quad f(x)<c<s\}$.
By \x{IND}, $A$ has  a supremum.
At this supremum, $f$ cannot be less than $s$ since $f$ is continuous on $[a,b]$ and $s$ is the supremum of $Ran(f)$.
\end{proof}

Throughout  this work $Id:\F\rightarrow \F$ denotes the identity function,
i.e.\ $Id:x\mapsto x$.

\begin{thm}[Mean Value Theorem]\label{thmLagrange}
Assume \x{IND}.
Let $f:\F\parrow \F$ be definable, differentiable on $[a,b]$.
If $a\neq b$, then there is an $s\in (a,b)$ such that  $f'(s)=\frac{f(b)-f(a)}{b-a}$.
\end{thm}
\begin{proof}
Assume $a\neq b$.
Let  $h:=\big(f(b)-f(a)\big)Id - (b-a)f$.
Then  $h$ is differentiable on $[a,b]$ and $h'(x)=f(b)-f(a)-(b-a)f'(x)$
for all $x\in [a,b]$ by
(ii) and (iii) of Proposition~\ref{propDiff} since  $Id$ is differentiable on $[a,b]$
and its derivate is $1$ for all $x\in [a,b]$.
If $h$ is constant on $[a,b]$,%
\footnote{ , i.e.\ if there is a $c\in \F$ such that $h(x)=c$ for all $x\in
[a,b]$,}
 then $h'(s)=0$ for all $s\in(a,b)$.
Otherwise, by Theorem~\ref{thmsup}, there is a maximum or minimum of $h$ different from
$h(a)=f(b)a-bf(a)=h(b)$ at an $s\in (a,b)$.
Hence $h'(s)=0$ by Proposition~\ref{propMax}.
This completes the proof since $a\neq b$ and  $h'(s)=f(b)-f(a)-(b-a)f'(s)$.\end{proof}

\begin{col}[Rolle's Theorem]\label{thmRoll}
Assume \x{IND}.
Let $f:\F\parrow \F$ be definable and  differentiable on $[a,b]$.
If $f(a)=f(b)$ and $a\neq b$, then there is an $s\in (a,b)$ such that  
$f'(s)=0$.
\phantom{iiiiii}\hfill\qedsymbol
\end{col}

\begin{prop} \label{propInt}
Assume \x{IND}.
Let  $f,g:\F\parrow \F$ be  definable and  differentiable on $(a,b)$.
If $f'(x)=g'(x)$ for all $x\in(a,b)$, then there is a $c\in \F$ such that 
$f(x)=g(x)+c$ for all $x\in(a,b)$.
\end{prop}

\begin{proof}
Assume that $f'(x)=g'(x)$ for all $x\in(a,b)$.
Let $h:=f-g$.
Then $h'(x)=f'(x)-g'(x)=0$ for all $x\in(a,b)$ by (ii) and (iii) of
Proposition~\ref{propDiff}.
If there are $y,z\in (a,b)$ such that $h(y)\neq h(z)$ and $y\neq z$, then, by the Mean Value
Theorem, there is an $x$ between $y$ and $z$ such that
$h'(x)=\frac{h(z)-h(y)}{z-y}\neq 0$ and this contradicts $h'(x)=0$.
Thus  $h(y)=h(z)$ for all $y,z\in (a,b)$. Hence there is a $c\in \F$ such that $h(x)=c$ for all $x\in(a,b)$.
\end{proof}

%\medskip
%%%%%%%%%%%%%%%%
\section{PROOFS OF THE MAIN RESULTS}%
\label{proofss}
%%%%%%%%%%%%%%%%
%\medskip

In the present section \x{AxFrame} is assumed without any further mentioning.

 Let $\widehat{\enskip}: \F \rightarrow \F^d$ be the natural
embedding defined as \ $\widehat{\enskip}: x \mapsto \langle
x,0,\dots,0\rangle$.
We define the {\bf life-curve} of observer $k$ as seen by observer
$m$ as $Tr^k_m:=\widehat{\enskip} \circ f^k_m$. Throughout  this
work we denote $\widehat{\enskip}(x)$ by $\widehat{x}$, for $x\in
\F$. Thus $Tr^k_m(t)$ is the coordinate point where  $m$  observes the 
event ``$k$'s
wristwatch shows $t$'', i.e. $Tr^k_m(t)=p$ iff
$ev_m(p)=ev_k(\langle t,0,\ldots,0\rangle)=ev_k(\,\widehat{t}\;)$.

In the following proposition, we list several easy but useful consequences of some of our axioms.

\begin{prop}\label{prop0}
Let $m\in \IOb$ and $k\in \Ob$.
Then (i)--(viii) below hold.
\begin{itemize}
\item[(i)] Assume \x{AxPh}.
Then $Cd(m)=\F^d$ and
for all distinct $p,q\in \F^d$,
$ev_m(p)\neq ev_m(q)$.
\item[(ii)] Assume \x{AxPh} and \x{AxSelf^-}. Then $tr_m(m)=\bar{t}$.
\item[(iii)] Assume \x{AxPh}.
Then $f^k_m:\F^d\parrow \F^d$ and $Tr^k_m:\F\parrow \F^d$.
\item[(iv)] Assume \x{AxPh} and \x{AxEv}. If $k\in \IOb$,
then $f^k_m:\F^d\rightarrow \F^d$ is
a bijection and $Tr^k_m:\F\rightarrow \F^d$ is an injection.
\item[(v)] Assume \x{AxEv}.
Let $h\in \IOb$. Then $f^k_h=f^k_m\circ f^m_h$ and
$Tr^k_h=Tr^k_m\circ f^m_h$.
\item[(vi)] Assume \x{AxAcc} and \x{AxEv}. Then $tr_k(k)\subseteq
Dom(f^k_m)$.
\item[(vii)] Assume \x{AxSelf^-}. Then
$\{\widehat{x}: x\in Dom(Tr^k_m)\}\subseteq tr_k(k)$ and
$Rng(Tr^k_m)\subseteq tr_m(k)$.
\item[(viii)] Assume \x{AxAcc}, \x{AxEv} and \x{AxSelf^-}.
Then $\{\widehat{x}: x\in Dom(Tr^k_m)\}=tr_k(k)$. 
\end{itemize}
\end{prop}

\begin{proof}
To prove (i), let $p,q\in \F^d$ be distinct points.
Then there is a line of slope $1$ that contains $p$ but does not contain $q$.
By \x{AxPh}, this line is  the trace of a photon.
For such a photon $ph$, we have $ph\in ev_m(p)$ and $ph\not\in ev_m(q)$.
Hence $ev_m(p)\neq ev_m(q)$ and $ev_m(p)\neq\emptyset$. Thus (i) holds.

\smallskip
\noindent
(ii) follows from (i) since $tr_m(m)=Cd(m)\cap \bar{t}$ by \x{AxSelf^-}.

\smallskip
\noindent
(iii) and (iv) follow from (i) by the definitions of the world-view transformation and the life-curve.

\smallskip
\noindent
To prove (v), let $\langle p,q\rangle \in f^k_h$.
Then $ev_k(p)=ev_h(q)\neq\emptyset$.
Since, by \x{AxEv}, $h$ and $m$ observe the same set of events, there is an $r\in
\F^d$ such that $ev_m(r)= ev_h(q)$.
But then $\langle p,r\rangle \in f^k_m$ and $\langle r,q\rangle \in f^m_h$.
Hence $\langle p,q\rangle \in f^k_m\circ f^m_h$.
Thus $f^k_h\subseteq f^k_m\circ f^m_h$.
The other inclusion follows from the definition of the world-view transformation.
Thus $f^k_h=f^k_m\circ f^m_h$ and $Tr^k_h=\widehat{\enskip}\circ f^k_h=\widehat{\enskip}\circ f^k_m\circ f^m_h=Tr^k_m\circ f^m_h$.

\smallskip
\noindent
To prove (vi), let $q\in tr_k(k)$.
By \x{AxAcc}, there is an $h\in \IOb$ such that $h$ is a co-moving observer of $k$
at $q$.
For such an $h$, we have $f^k_h(q)=q$ and, by (v), $Dom (f^k_h)\subseteq Dom (f^k_m)$.
Thus $q\in Dom (f^k_m)$.

\smallskip
\noindent
To prove (vii), let $\langle x,q\rangle \in Tr^k_m$. Then
$\langle\widehat{x},q\rangle\in f^k_m$.
But then $ev_k(\widehat{x})=ev_m(q)\neq\emptyset$.
Thus $\widehat{x}\in Cd(k)\cap\bar{t}$. By \x{AxSelf^-}, $\widehat{x}\in tr_k(k)$; and this proves the first part of (vii).
By $\widehat{x}\in tr_k(k)$, we have $k\in ev_k(\widehat{x})=ev_m(q)$.
Thus $q\in tr_m(k)$ and this proves the second part of (vii).

\smallskip
\noindent
The ``$\subseteq$ part'' of (viii) follows from (vii).
To prove the other inclusion, let $p\in tr_k(k)$.
Then, by \x{AxSelf^-} and (vi), $p\in\bar t\cap Dom (f^k_m)$.
Thus there are $x\in \F$ and $q\in \F^d$ such that $\widehat{x}=p$ and $\langle
p,q\rangle\in f^k_m$.
But then $\langle x,q\rangle \in\widehat{\enskip}\circ f^k_m= Tr^k_m$.
Hence $x\in Dom(Tr^k_m)$.
\end{proof}

We say that $f$  is {\bf well-parametrized}  iff
$f:\F\parrow \F^d$ and the following holds:
if $x\in Dom(f)$ is an accumulation point of $Dom(f)$,
then $f$ is differentiable at $x$ and its derivate at $x$ is of
Minkowski-length $1$, i.e.\ $\mu\big(f'(x)\big)=1$.
Assume $\mathfrak{F}=\mathfrak{R}$. Then the 
curve  $f$  is well-parametrized iff  $f$  is
parametrized according to Minkowski-length, i.e.\ for all $x,y\in \F$,
if $[x,y]\subseteq Dom(f)$,  
the Minkowski-length of  $f$ restricted  to $[x,y]$ is  $y-x$. 
(By
Minkowski-length of a curve we mean length according to
Minkowski-metric, e.g., in the sense of Wald~\cite[p.43, (3.3.7)]{Wal}).
{\bf Proper time} or {\bf wristwatch time} is defined as the Minkowski-length
of a time-like curve, cf.\ e.g., Wald~\cite[p.44, (3.3.8)]{Wal}, 
Taylor-Wheeler~\cite[1-1-2]{TW00} 
 or d'Inverno~\cite[p.112, (8.14)]{d'Inverno}. 
Thus a curve defined on
a subset of $\mathfrak{R}$ 
is well-parametrized iff it is parametrized according
to proper time, or wristwatch-time. (Cf.\ e.g.,
\cite[p.112, (8.16)]{d'Inverno}.)

The next proposition states that life-curves of accelerated observers in
models of \x{AccRel_0} are well-parametrized.
This implies that
accelerated clocks behave as expected in models of \x{AccRel_0}.  
Remark~\ref{Trrem} after the
proposition will state a kind of ``completeness theorem'' for life-curves
of accelerated observers, much in the spirit of  Remark~\ref{rem-specthm}.

\begin{prop}\label{propTr}
Assume \x{AccRel_0} and $d>2$. Let $m\in \IOb$ and $k\in \Ob$. Then $Tr^k_m$
is well-parametrized and definable.
\end{prop}

\begin{proof}
Let $m\in \IOb$, $k\in \Ob$. Then $Tr^k_m$ is definable by  its
definition.
Furthermore, $f^k_m:\F^d\parrow \F^d$ and $Tr^k_m:
\F\parrow \F^d$ by (iii) of Proposition~\ref{prop0}.
 Let $x\in Dom(Tr^k_m)$ be an accumulation point of $Dom(Tr^k_m)$.
We would like to prove that $Tr^k_m$ is differentiable at $x$ and
its derivate at $x$ is of Minkowski-length $1$.
$\widehat{x}\in tr_k(k)$ by (vii) of Proposition~\ref{prop0}.
Thus, by \x{AxAcc}, there is a co-moving inertial observer of $k$
at $\widehat{x}$. By Proposition~\ref{propAff}, we
can assume that $m$ is a co-moving inertial observer of
$k$ at $\widehat{x}$, i.e.\ $m\succ_{\widehat{x}}k$,
 because of the following three statements.
By (v) of Proposition~\ref{prop0}, for every $h\in \IOb$,
either of $Tr^k_m$ and $Tr^k_h$ can be obtained from the other by
composing the other by a world-view transformation between inertial observers.
By Theorem~\ref{thmPoi}, world-view transformations between inertial observers are Poincar\'e-transformations.
Poincar\'e-transformations are affine and preserve the Minkowski-distance.

Now, assume that $m$ is a co-moving inertial observer of
$k$ at $\widehat{x}$. Then $f^k_m(\widehat{x})=\widehat{x}$,
 $z\widehat{1}=\widehat{z}$ and $Tr^k_m(z)=f^k_m(\widehat{z})$  for
every $z\in Dom(Tr^k_m)$.
Therefore
\begin{equation}
\forall y\in Dom(Tr^k_m)\quad
\big|Tr^k_m(y)-Tr^k_m(x)-(y-x)\widehat{1}\big|=\big|f^k_m(\widehat{y})-\widehat{y}\big|.
\label{tp-e1}
\end{equation}
Since $Dom(f^k_m)\subseteq Cd(k)$ and $\widehat{y}\in Dom(f^k_m)$ if $y\in Dom
(Tr^k_m)$, we have that for all $\delta\in \F^+$,
\begin{equation}
\forall y\in Dom(Tr^k_m)\quad |y-x|<\delta\; \Longrightarrow\; \widehat{y}\in
B_\delta(\widehat{x})\cap Cd(k).
\label{tp-e2}
\end{equation}
Let $\varepsilon \in \F^+$ be fixed. Since
$m\succ_{\widehat{x}} k$ and $f^k_m: \F^d\parrow \F^d$, there is a $\delta\in
\F^+$
such that
\begin{equation}
\forall p\in B_\delta(\widehat{x})\cap Cd(k)\quad \big|p-f^k_m(p)\big|\leq \varepsilon
|p-\widehat{x}|.
\label{tp-e3}
\end{equation}
Let such a $\delta$ be fixed.
By (\ref{tp-e2}), (\ref{tp-e3}) and the fact that
$|\widehat{y}-\widehat{x}|=|y-x|$, we have that
\begin{equation}
\forall y\in Dom(Tr^k_m)\quad |y-x|<\delta\; \Longrightarrow\;
\big|\widehat{y}-f^k_m(\widehat{y})\big|\leq \varepsilon |y-x|.
\end{equation}
By this and (\ref{tp-e1}), we have
\begin{equation}
\begin{split}
\forall y\in Dom(Tr^k_m) \quad  & |y-x|<\delta \; \Longrightarrow\\
&\big|Tr^k_m(y)-Tr^k_m(x)-(y-x)\widehat{1}\big|\leq\varepsilon |y-x|.
\end{split}
\end{equation}
Thus $(Tr^k_m)'(x)=\widehat{1}$. This completes the proof
since $\mu(\,\widehat{1}\,)=1$.
\end{proof}

\begin{rem}
\label{Trrem} Well parametrized curves are exactly the life-curves of
accelerated observers, in models of \x{AccRel_0}, as follows.
Let $\mathfrak{F}$ be an Euclidean ordered field and
let $f:\F\parrow \F^d$ be well-parametrized. Then there are a model
$\mathfrak{M}$ of \x{AccRel_0}, $m\in \IOb$ and $k\in \Ob$ such that
$Tr^k_m=f$ and the ordered field reduct of $\mathfrak{M}$ is
$\mathfrak{F}$. Recall that if $\mathfrak{F}=\mathfrak{R}$, then
this $\mathfrak{M}$ is a model of \x{AccRel}. This is not difficult to prove
by using the methods of the present paper.
\phantom{iiiiii}\vege 
\end{rem}

We say that $p\in \F^d$ is {\bf vertical} iff $p\in \bar{t}$.

\begin{lem} \label{lemWp}
Let $f:\F\parrow \F^d$ be well-parametrized. Then (i) and (ii) below hold.
\begin{itemize}
\item[(i)] Let  $x\in Dom(f)$ be an accumulation point of $Dom(f)$.
Then $f_t$ is differentiable at $x$ and $|f'_t(x)|\ge
1$. Furthermore, $|f'_t(x)|=1$ iff $f'(x)$  is vertical.
\item[(ii)] Assume \x{IND} and that $f$ is
definable. Let $[a,b]\subseteq Dom(f)$. Then $f_t$ is increasing or decreasing on $[a,b]$.
If $f_t$ is increasing on $[a,b]$ and $a\neq b$, then 
$f'_t(x)\ge 1$ for all $x\in [a,b]$. 
\end{itemize}
\end{lem}

\begin{proof}
Let $f$ be well-parametrized.

\smallskip
\noindent
To prove (i), let $x\in Dom(f)$ be an accumulation point of $Dom(f)$.
Then $f'(x)$ is of Minkowski-length $1$.
By Proposition~\ref{propAff}, $f_t$ is differentiable at $x$ and $f'_t(x)=f'(x)_t$.
Now, (i) follows from the fact that the absolute value of the time component of a
vector of Minkowski-length 1 is always greater than 1 and it is 1 iff the vector is vertical.

\smallskip
\noindent
To prove (ii), assume \x{IND} and that $f$ is definable.
Let $[a,b]\subseteq Dom(f)$. From (i), we have $f'_t(x)\neq 0$ for all $x\in[a,b]$.
Thus, by Rolle's theorem, $f_t$ is injective on $[a,b]$.
Thus, by Bolzano's theorem, $f_t$ is increasing or decreasing on $[a,b]$
since $f_t$ is continuous and injective on $[a,b]$. Assume that
$f_t$ is increasing on $[a,b]$ and $a\neq b$. Then $f'_t(x)\geq 0$ for all $x\in [a,b]$  by the definition of the derivate.
Hence, by (i), $f'_t(x)\geq 1$ for all $x\in [a,b]$.
\end{proof}

\goodbreak

\begin{thm}\label{thmJtwp}
Assume \x{IND}.
Let $f:\F\parrow \F^d$ be definable, well-parametrized and
$[a,b]\subseteq Dom(f)$.
Then (i) and (ii) below hold.
\begin{itemize}
\item[(i)] $b-a\le \left|f_t(b)-f_t(a)\right|$.
\item[(ii)] If $f(x)_s\neq f(a)_s$ for an $x\in[a,b]$, then
$b-a<\big|f_t(b)-f_t(a)\big|$. 
\end{itemize}
\end{thm}

\begin{proof}
Let $f:\F\parrow \F^d$ be definable, well-parametrized and
$[a,b]\subseteq Dom(f)$. We can assume that $a\neq b$.
For every $i\leq d$, $f_i$ is definable and differentiable on $[a,b]$ by Proposition~\ref{propAff}.
Then, by the Main Value Theorem,
there is an $s\in (a,b)$ such that $f'_t(s)=\frac{f_t(b)-f_t(a)}{b-a}$.
By (i) of Lemma~\ref{lemWp}, we have $1\le |f'_t(s)|$.
But then, $b-a\le \big|f_t(b)-f_t(a)\big|$. This completes the proof of (i).

\smallskip
\noindent
To prove  (ii), 
let $x\in [a,b]$ be such that $f(x)_s\neq f(a)_s$.
Let $1<i\le d$ be such that $f_i(x)\neq f_i(a)$.
Then, by the Main Value Theorem, there is an
$y\in (a,b)$ such that $f'_i(y)=\frac{f_i(x)-f_i(a)}{x-a}\neq 0$.
Thus $f'(y)$ is not vertical.
Therefore, by (i) of  Lemma~\ref{lemWp}, we have $1<|f'_t(y)|$.
Thus, by the definition of the derivate,
there is a $z\in(y,b)$ such that $1<\frac{|f_t(z)-f_t(y)|}{z-y}$.
Hence we have
\begin{equation}
z-y<|f_t(z)-f_t(y)|.
\end{equation}
Let us note that $a<y<z<b$. By applying (i) 
to $[a,y]$ and $[z,b]$, respectively, we get
\begin{equation}
y-a\le \big| f_t(y)-f_t(a) \big|\quad\text{and}\quad b-z\le \big|f_t(b)-f_t(z)\big|.
\end{equation}
$f_t$ is increasing or decreasing on $[a,b]$  by (ii) of Lemma~\ref{lemWp}.
Thus $f_t(a)<f_t(y)<f_t(z)<f_t(b)$ or $f_t(a)>f_t(y)>f_t(z)>f_t(b)$.
Now, by adding up the last three inequalities, we get $b-a<\big|f_t(b)-f_t(a)\big|$.
\end{proof}

Let $a\in \F^d$. For convenience, we introduce the following notation:
$a^+:=a$  if  $a_t \ge 0$ and $a^+:=-a$ if $a_t < 0$.
A set $H\subseteq \F^d$ is called {\bf twin-paradoxical} iff $\widehat{1}\in H$,
$o\not\in H$, $\text{\it slope}(p)<1$ if $p\in H$,
for all $p\in \F^d$ if  $\text{\it slope}(p)<1$,
then there is a $\lambda\in \F$ such that $\lambda p \in H$, and
for all distinct $p,q,r\in H$ and for all $\lambda,\mu \in \F^+$, $r^+=\lambda p^+ + \mu q^+$
implies that  $\lambda+\mu<1$.

A positive answer to the following question would also provide  a positive
answer to Question~\ref{qTwp}, cf.\  \cite[\S 3]{mythes}.

\begin{que}
\label{qConv}
Assume \x{IND}.
Let $f:\F\parrow \F^d$ be  definable such that $f$ is differentiable on
$[a,b]$ and $f(a),f(b)\in\bar{t}$.
Furthermore, let the set $\{f'(x):x\in [a,b] \}$ be a subset of a twin-paradoxical set.
Are then (i) and (ii) below true?
\begin{itemize}
\item[(i)] $b-a\le \big|f_t(b)-f_t(a)\big|$.
\item[(ii)] If $f(x)_s\neq f(a)_s$ for an $x\in[a,b]$, then
$b-a<\big|f_t(b)-f_t(a)\big|$. \vege
\end{itemize}
\end{que}

\begin{thm}\label{thmJeq}
Assume \x{IND}.
Let $f,g:\F\parrow \F^d$ be definable and well-parametrized.
Let $[a,b]\subseteq Dom(f)$ and $[a',b']\subseteq Dom(g)$
be such that $\{f(r):r\in[a,b]\}=\{g(r'):r'\in [a',b']\}$. Then $b-a=b'-a'$.
\end{thm}

\begin{proof}
By (ii) of Lemma~\ref{lemWp},
$f_t$ is increasing or decreasing on $[a,b]$ and
so is $g_t$ on $[a',b']$. We can assume that $Dom(f)=[a,b]$,
$Dom(g)=[a',b']$
and that $f_t$ and $g_t$ are increasing on $[a,b]$ and $[a',b']$,
respectively.%
\footnote{ It can be assumed that $f_t$ is increasing on $[a,b]$ because
the assumptions of the theorem remain true when $f$ and $[a,b]$ are replaced by
$-Id\circ f$ and $[-b,-a]$, respectively, and $f_t$ is
decreasing on $[a,b]$ iff
$(-Id\circ f)_t$ is increasing on $[-b,-a]$.}
Then $Rng(f)=Rng(g)$.
Furthermore, $f$ and $g$ are injective since $f_t$ and $g_t$ are such.
Since $Rng(f)=Rng(g)$ and $g_t$ is injective,
$f\circ g^{-1}=f_t\circ g_t^{-1}$.
 Let $h:=f\circ g^{-1}=f_t\circ g_t^{-1}$.
Since $Rng(f_t)=Rng(g_t)$
and $f_t$ and $g_t$ are increasing, $h$ is an increasing bijection between $[a,b]$ and $[a',b']$.
Hence $h(a)=a'$ and $h(b)=b'$.
We are going to prove that $b-a=b'-a'$ by proving that there is a $c\in \F$ such that $h(x)=x+c$ for all $x\in [a,b]$.
We can assume that $a\neq b$ and $a'\neq b'$.
By Lemma ~\ref{lemWp}, $f_t$ and $g_t$ are differentiable on $[a,b]$ and
$[a',b']$, respectively, and $f'_t(x)>0$ for all $x\in[a,b]$ and  $g'_t(x')>0$ for all
$x'\in[a',b']$.
By (iv) and (v) of Proposition~\ref{propDiff}, $h=f_t\circ g_t^{-1}$ is
also differentiable on $(a,b)$.
By $h=f\circ g^{-1}$, we have $f=h\circ g$.
Thus  $f'(x)=h'(x)g'\big(h(x)\big)$
for all $x\in(a,b)$ by (iv) of Proposition~\ref{propDiff}.
Since both  $f'(x)$ and $g'\big(h(x)\big)$ are of Minkowski-length $1$ and
their time-components are positive%
\footnote{ i.e.\ $f'_t(x)>0$ and
$g'_t\big(h(x)\big)>0$}
for all $x\in(a,b)$, we conclude  that $h'(x)=1$ for all $x\in(a,b)$.
By Proposition~\ref{propInt}, we get that there is a $c\in \F$ such
that $h(x)=x+c$  for all $x\in(a,b)$  and thus for all $x\in[a,b]$ since $h$ is
an increasing bijection
between $[a,b]$ and $[a',b']$.
\end{proof}

\begin{proof}[\textcolor{proofcolor}{proof of Theorem~\ref{thmTwp}}]
\label{thmTwp-proof}
Assume \x{AccRel} and $d>2$.
Let $m\in \IOb$ and $k\in \Ob$.
Let $p,q\in tr_k(k)$, $p',q'\in tr_m(m)$ be such that $\langle
p,p'\rangle,\langle q,q'\rangle \in f^k_m$, $[p q]\subseteq tr_k(k)$
and $[p' q']\not\subseteq tr_m(k)$, cf.\ Figure~2.
Let us abbreviate $Tr^k_m$ by $Tr$.
We are going to prove that $|q_t-p_t|<|q'_t-p'_t|$ by applying Theorem~\ref{thmJtwp} to $Tr$ and
$[p_t, q_t]$. By Proposition~\ref{propTr},
\begin{equation}
\label{twp-e1}
Tr: \F\parrow \F^d \text{ is
well-parametrized and definable.}
\end{equation}
By \x{AxSelf^-}, $p,q,p',q'\in\bar{t}$.
By $\widehat{p_t}=p$, by $\widehat{q_t}=q$, by $\langle p,p'\rangle ,\langle
q,q'\rangle \in f^k_m$ and by $Tr=\widehat{\enskip}\circ f^k_m$,
\begin{equation}
\label{twp-e2}
Tr(p_t)=p'\quad\text{ and }\quad Tr(q_t)=q'.
\end{equation}
By $p,q\in tr_k(k)$ and $\langle p,p'\rangle, \langle q,q'\rangle\in f^k_m$,
we have that $p',q'\in tr_m(k)$. Thus, by
$[p' q']\not\subseteq tr_m(k)$, we have that $p'\neq q'$. Hence, by
(\ref{twp-e2}), $p_t\neq q_t$. We can assume that $p_t<q_t$.
By (viii) of Proposition~\ref{prop0}, $\{\widehat{x}: x\in Dom(Tr)\}=tr_k(k)$.
Since $[p q]\subseteq tr_k(k)$,
\begin{equation}
\label{twp-e3}
[p_t, q_t]\subseteq Dom(Tr).
\end{equation}
By (i) of Lemma~\ref{lemWp}, (\ref{twp-e1}) and (\ref{twp-e3}), 
we have that $Tr_t$ is differentiable on
$[p_t,q_t]$, thus it is continuous on $[p_t,q_t]$.
Let $x'\in [p' q']\subseteq \bar{t}$ be such that $x'\not\in tr_m(k)$.
By Bolzano's theorem and (\ref{twp-e2}), there is an
$x\in [p_t,q_t]$
such that $Tr_t(x)=x'_t$. Let such an $x$ be fixed.
$Tr(x)\in tr_m(k)$ since $Rng(Tr)\subseteq tr_m(k)$ by (vii) of
Proposition~\ref{prop0}.
But then $Tr(x)\neq x'$.
Hence $Tr(x)\not\in\bar{t}$. Thus
\begin{equation}
\label{twp-e4}
x\in [p_t,q_t]\quad\mbox{and}\quad Tr(x)_s\neq Tr(p_t)_s
\end{equation}
since $Tr(p_t)=p'\in\bar t$.
Now, by (\ref{twp-e1})--(\ref{twp-e4}) above,
we can  apply (ii) of Theorem~\ref{thmJtwp} to $Tr$ and $[p_t,q_t]$,
and we get that $|q_t-p_t|<|Tr_t(q_t)-Tr_t(p_t)|=|q'_t-p'_t|$.
\end{proof}

\begin{proof}[\textcolor{proofcolor}{proof of Theorem~\ref{thmEq}}]
\label{thmEq-proof}
Assume \x{AccRel} and $d>2$.
Let $k$ and $m$ be observers.
Let $p,q \in tr_k(k)$, $p',q'\in tr_m(m)$ be such that
$\emptyset\not\in\{ev_k(r):r\in[p q]\}=\{ev_m(r'):r'\in [p' q']\}$, cf.\ the
right hand side of Figure~2. 
Thus $[p q]\subseteq Cd(k)$ and $[p' q']\subseteq Cd(m)$.
By \x{AxSelf^-}, $ tr_k(k)= Cd(k)\cap \bar t$ and
$tr_m(m)= Cd(m)\cap \bar t$. Therefore $[p q]\subseteq tr_k(k)\subseteq\bar
t$ and $[p' q']\subseteq tr_m(m)\subseteq\bar t$. We can assume that
$p_t\leq q_t$ and $p'_t\leq q'_t$.
Let $h\in \IOb$.
We are going to prove that
$\big|q_t-p_t\big|=\big|q'_t-p'_t\big|$, by applying Theorem~\ref{thmJeq} as follows:
let
$[a,b]:=[p_t,q_t]$, $[a',b']:=[p'_t,q'_t]$, $f:=Tr^k_h$ and $g:=Tr^m_h$.
By (viii) of Proposition~\ref{prop0}, by $[p q]\subseteq tr_k(k)$
and by $[p' q']\subseteq tr_m(m)$, we conclude that $[a,b]\subseteq Dom(f)$ and $[a',b']\subseteq Dom(g)$.
By Proposition~\ref{propTr}, $f$ and $g$ are well-parametrized and definable.
We have $\{f(r):r\in[a,b]\}=\{g(r'):r'\in [a',b']\}$ since $\{ev_k(r):r\in[p q]\}=\{ev_m(r'):r'\in [p' q']\}$.
Thus, by Theorem~\ref{thmJeq}, we conclude that $b-a=b'-a'$.
Thus $\big|q_t-p_t\big|=\big|q'_t-p'_t\big|$ and this is what we wanted to prove.
\end{proof}

\begin{proof}[\textcolor{proofcolor}{proofs of Theorems~\ref{thmNoIND} and \ref{thmMO}}]
\label{thmNoIND-proof}
\label{thmMO-proof}
We will construct three models. 
Let $\mathfrak{F}=\left<\F;+,\cdot,\le \right>$ be an  Euclidean ordered field
different from $\mathfrak{R}$.
For every $p\in \F^d$, let $m_p:\F^d\rightarrow \F^d$ denote the translation 
by vector $p$, i.e.\ $m_p: q\mapsto q+p$.
$f:\F^d\rightarrow \F^d$ is called  {\bf translation-like} if{}f for all 
$q \in\F^d$,  there is
a  $\delta \in \F^+$ such that for all  $p\in B_\delta(q)$,
$f(p)=m_{f(q)-q}(p)$ and for all $p,q\in \F^d$, $f(p)=f(q)$ and $p\in
\bar t$ imply that $q\in\bar t$. Let $k:\F^d\rightarrow \F^d$ be
translation-like.
First we construct a model $\mathfrak{M}_{(\mathfrak{F},k)}$  of \x{AccRel_0} and (i) and (ii) of Theorem~\ref{thmMO}
for $\mathfrak{F}$ and $k$, which will be a model of (iii) and (iv) of
Theorem~\ref{thmMO} if $k$ is a bijection.
We will show that \x{Tp} is false in $\mathfrak{M}_{(\mathfrak{F},k)}$.
Then we will choose $\mathfrak{F}$ and $k$ appropriately to get the desired models
in which \x{Ddpe} is false, too.

Let the ordered field reduct of $\mathfrak{M}_{(\mathfrak{F},k)}$ be
$\mathfrak{F}$. Let $\{I_1, I_2, I_3, I_4, I_5\}$ be a partition%
\footnote{ I.e.\ $I_i$'s are disjoint and $\F=I_1\cup I_2 \cup I_3 \cup 
I_4 \cup I_5$.}
 of $\F$ such that every $I_i$ is open, $x\in I_2 \iff x+1\in I_3 \iff x+2 \in I_4$ and for all $y\in I_i$ and
$z\in I_j$, $y\leq z \iff i\leq j$.
Such a partition can easily be constructed.%
\footnote{ Let $H\subset \F$ be a non-empty bounded set that does not have a supremum. Let $I_1:=\{x\in
\F: \exists h \in H \quad x<h\}$, $I_2:=\{x+1\in \F: x\in I_1\}\setminus I_1$, $I_3:=\{x+1\in
\F: x\in I_2\}$,  $I_4:=\{x+1\in \F: x\in I_3\}$ and $I_5:=\F\setminus(I_1\cup I_2 \cup I_3 \cup I_4)$.}
Let
\begin{equation}
k'(p):=\left\{
\begin{array}{cl}
  p & \text{ if } p_t\in I_1\cup I_5 , \\
p-\widehat{1} & \text{ if }  p_t\in I_4 ,  \\
p+\widehat{1} & \text{ if } p_t\in I_3 ,  \\
p+\langle 0,1,0,\ldots,0\rangle & \text{ if } p_t \in I_2
 \end{array}
\right.
\end{equation}
for every $p\in \F^d$, cf.\ Figure~4.
\begin{figure}
\small
\begin{center}
\psfrag*{p'}[r][r]{$p'$}
\psfrag*{q'}[r][r]{$q'$}
\psfrag*{q}[r][r]{$q$}
\psfrag*{p}[r][r]{$p$}
\psfrag*{text1}[b][b]{world-view of $m$}
\psfrag*{text2}[b][b]{world-view of $k'$}
\psfrag*{I0}[l][l]{$I_1$}
\psfrag*{I1}[l][l]{$I_2$}
\psfrag*{I2}[l][l]{$I_3$}
\psfrag*{I3}[l][l]{$I_4$}
\psfrag*{I4}[l][l]{$I_5$}
\psfrag*{tr}[r][r]{$tr_{k'}(k')$}
\psfrag*{tr1}[r][r]{$tr_m(k')$}
\psfrag*{k}[b][b]{$k'$}
\psfrag*{f}[t][t]{$f^{k'}_m$}
\psfrag*{ff}[t][t]{$f^m_{k'}$}
\psfrag*{text3}[t][t]{\shortstack[c]{$\mathfrak{F}\neq\mathfrak{
R}$ is
Euclidean,\\
\x{Tp} is false:\\
$|p_t-q_t|>|p'_t-q'_t|$}}
\includegraphics[keepaspectratio, width=0.9\textwidth]{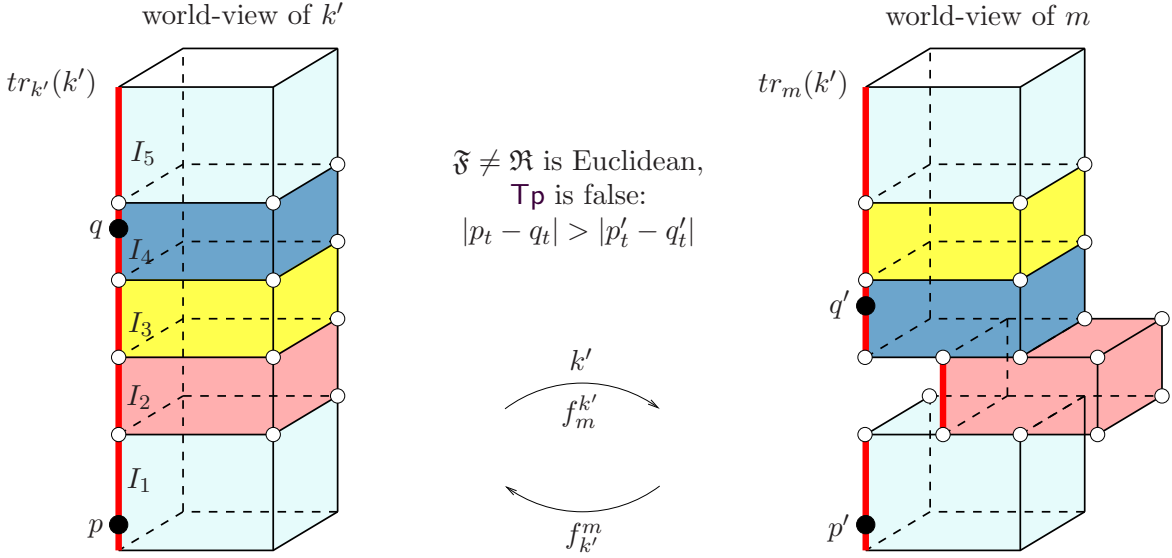}
\caption{\label{notwp} for the proofs of Theorems~\ref{thmNoIND} and \ref{thmMO}.}
\end{center}
\end{figure}
It is easy to see that $k'$ is  a translation-like bijection.
Let $\IOb:=\{m_p:p\in \F^d\}$, $\Ob:=\IOb\cup\{k, k'\}$, $\Ph:=\{l\in \text{\it Lines}: \text{\it slope}(l)=1\}$ and
$\B:=\Ob\cup \Ph$. Recall that $o:=\langle 0,\ldots,0\rangle$ is the origin.
First we give the world-view of $m_o$ then we give the world-view of an arbitrary observer $h$ by 
giving the world-view
transformation between $h$ and $m_o$. Let $ tr_{m_o}(ph):=ph$ and
$tr_{m_o}(h):=\{h(x): x\in\bar{t}\,\}$ for all $ph\in \Ph$ and $h\in \Ob$.
And let $ev_{m_o}(p):=\{b\in \B: p\in tr_{m_o}(b)\}$
for all $p\in \F^d$.
Let $f^h_{m_o}:=h$ for all $h\in \Ob$.
From these world-view transformations, we can obtain the world-view of each observer
$h$ in the following way: $ev_h(p):=ev_{m_o}\big(h(p)\big)$ for all $p\in
\F^d$. And from the world-views, we can obtain the $\W$ relation as follows:
for all $h\in \Ob$, $b\in \B$ and $p\in \F^d$, let $\W(h,b,p)$ iff $b\in ev_h(p)$.
Thus we are given the model  $\mathfrak{M}_{(\mathfrak{F},k)}$.
We note that $f^m_h=m\circ h^{-1}$ and $m_{h(q)-q}\succ_q h$ for all $m,h\in
\Ob$
and $q\in \F^d$.
It is easy to check that the axioms of \x{AccRel_0} and (i) and (ii) of
Theorem~\ref{thmMO} are true in  $\mathfrak{M}_{(\mathfrak{F},k)}$ and that if $k$ is a bijection, then  (iii) and (iv) of
Theorem~\ref{thmMO} are also true in $\mathfrak{M}_{(\mathfrak{F},k)}$.
Let $p,q\in\bar t$ be such that $p_t\in I_1$, $q_t\in I_4$;
and let $p':=k'(p)=p$, $q':=k'(q)=q-\widehat{1}$ and  $m:=m_{o}$.
It is  easy to check that \x{Tp} is false in $\mathfrak{M}_{(\mathfrak{F},k)}$ for
$k'$, $m$, $p$, $q$, $p'$ and $q'$, i.e.
$p,q \in tr_{k'}(k')$, $p',q' \in tr_m(m)$,
$\langle p,p'\rangle ,\langle q,q'\rangle \in f^{k'}_m$,
$[p q] \subseteq tr_{k'}(k')$, $[p' q']\not\subseteq tr_m(k')$
and $\big|q_t-p_t\big|\not<\big|q'_t-p'_t\big|$,
cf.\ Figure~4.

\begin{figure}
\begin{center}
\small
\psfrag*{text2}[t][t]{world-view of $m$}
\psfrag*{text1}[t][t]{world-view of $k$}
\psfrag*{p'}[r][r]{$p'$}
\psfrag*{q'}[r][r]{$q'$}
\psfrag*{q}[r][r]{$q$}
\psfrag*{p}[r][r]{$p$}
\psfrag*{P}[rt][rt]{$p$}
\psfrag*{Q}[lb][lb]{$q$}
\psfrag*{P'}[rt][rt]{$p'$}
\psfrag*{Q'}[lb][lb]{$q'$}
\psfrag*{tr}[r][r]{$tr_k(k)$}
\psfrag*{tr1}[r][r]{$tr_m(k)$}
\psfrag*{k}[b][b]{$k$}
\psfrag*{f}[t][t]{$f^k_m$}
\psfrag*{ff}[t][t]{$f^m_k$}
\psfrag*{text6}[t][t]{\shortstack[c]{first model, \\$\mathfrak{F}\neq\mathfrak{R}$ is
Euclidean,\\
\x{Ddpe} is false:\\
$|p_t-q_t|\neq|p'_t-q'_t|$}}
\psfrag*{text4}[t][t]{\shortstack[c]{second model,\\ $\mathfrak{F}$ is
non-Archimedean,\\ \x{Ddpe} is false:\\
$|p_t-q_t|\neq |p'_t-q'_t|$}}
\psfrag*{text5}[t][t]{\shortstack{third model,\\ $\mathfrak{F}$ is countable
Archimedean,\\ \x{Ddpe} is false:\\
$|p_t-q_t|\neq |p'_t-q'_t|$}}
\psfrag*{tr2}[l][l]{$tr_m(m)$}
\psfrag*{tr0}[r][r]{$tr_m(m)$}
\psfrag*{a+1}[r][r]{$a+1$}
\psfrag*{a+2}[r][r]{$a+2$}
\psfrag*{a}[r][r]{$a$}
\psfrag*{1t}[l][l]{$\widehat{1}$}
\psfrag*{o}[l][l]{$o$}
\psfrag*{I1}[l][l]{$I_1$}
\psfrag*{I2}[l][l]{$I_2$}
\includegraphics[keepaspectratio, width=0.85\textwidth]{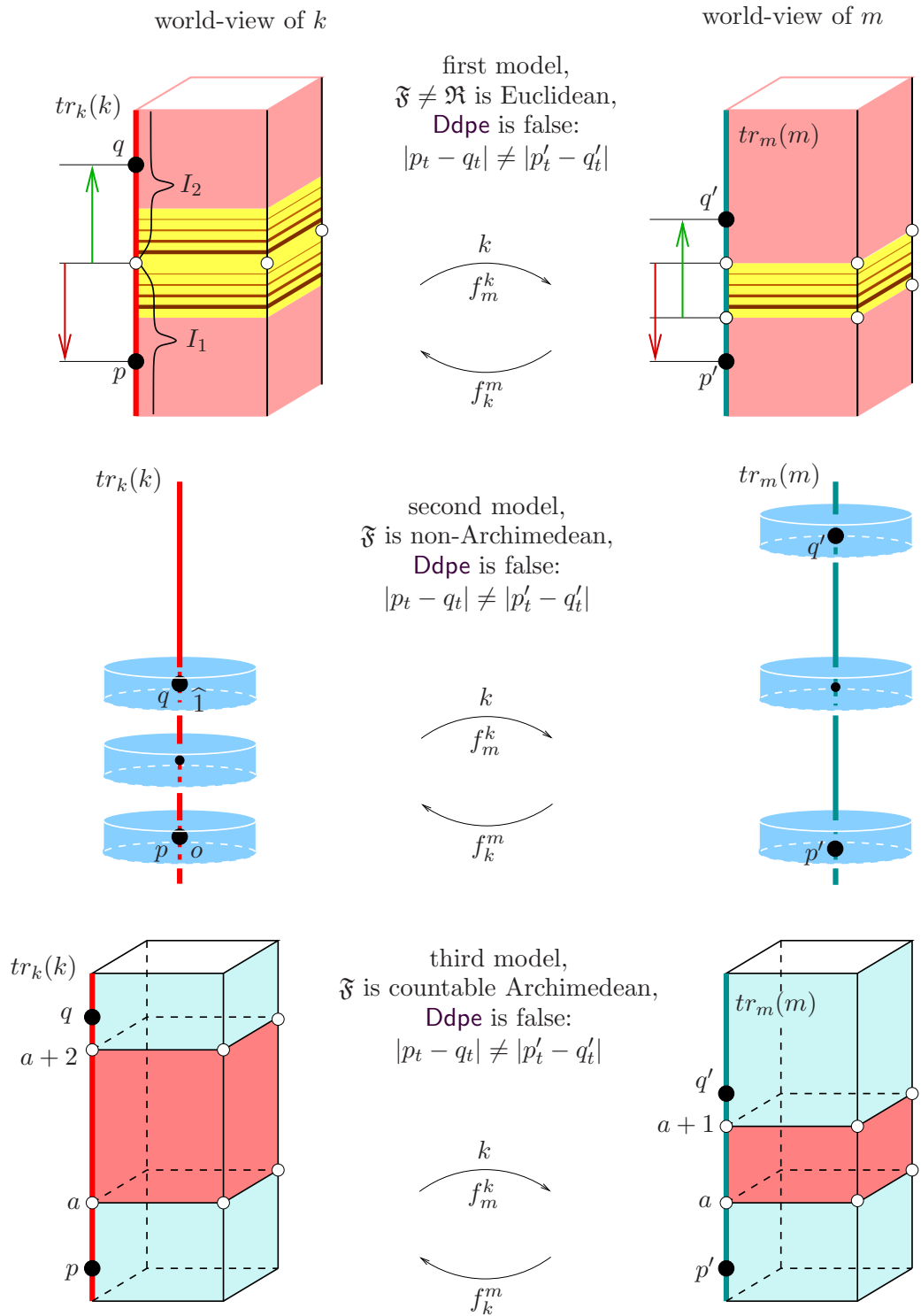}
\caption{\label{noDDPE} for the proofs of Theorems~\ref{thmNoIND} and \ref{thmMO}.}
\end{center}
\end{figure}

To construct the first  model, 
let $\mathfrak{F}$ be an arbitrary Euclidean ordered field different from 
$\mathfrak{R}$ and let $\{I_1, I_2\}$ be a partition of
$\F$ such that
for all $x\in I_1$ and $y\in I_2$, $x<y$. Let
\begin{equation}
k(p):=\left\{
\begin{array}{cl}
  p & \text{ if } p_t\in I_1 , \\
p-\widehat{1} & \text{ if }  p_t\in I_2
\end{array}
\right.
\end{equation}
for every $p\in \F^d$, cf.\ Figure~5.
It is easy to see that $k$ is translation-like.
Let $p,q\in\bar t$ be such that
$p_t,p_t+1\in I_1$ and $q_t,q_t-1\in I_2$; and let $p':=k(p)=p$,
$q':=k(q)=q-\widehat{1}$ and
$m:=m_{o}$. It is also easy to check that \x{Ddpe} is false in $\mathfrak{M}_{(\mathfrak{F},k)}$
for $k$, $m$, $p$, $q$, $p'$ and $q'$, i.e.\
$p,q \in tr_k(k)$, $p',q'\in tr_m(m)$,
$\emptyset\not\in\{ev_k(r):r\in[p q]\}=\{ev_m(r'):r'\in [p' q']\}$
and $\big|q_t-p_t\big|\neq\big|q'_t-p'_t\big|$, cf.\ Figure~5.
This completes the proof of Theorem~\ref{thmNoIND}.

To construct the second  model, let $\mathfrak{F}$ be an arbitrary non-Archimedean Euclidean ordered field.
Let $a\sim b$ if $a,b\in \F$ and  $a-b$ is infinitesimally small.
It is easy to see that $\sim$ is an equivalence relation.
Let us choose an element from every equivalence class of $\sim$ and let
$\tilde{a}$ denote the chosen element equivalent with $a\in \F$.
Let $k(p):=\langle p_t+\tilde{p}_t,p_s\rangle$ for every $p\in \F^d$, cf.\
Figure~5.
It is easy to see that $k$ is a translation-like bijection.
Let $p:=o$, $q:=\widehat{1}$, $p':=k(p)=\langle \tilde{0},0,\ldots,0\rangle$,
$q':=k(q)=\langle 1+\tilde{1},0,\dots,0\rangle$ and
$m:=m_{o}$.
It is also easy to check that \x{Ddpe} is false in $\mathfrak{M}_{(\mathfrak{F},k)}$  for $k$, $m$, $p$, $q$, $p'$ and
$q'$,  cf.\ Figure~5.

\begin{figure}[h!btp]
\begin{center}
\small
\psfrag*{a}[t][t]{$a$}
\psfrag*{r1}[t][t]{$r_1$}
\psfrag*{r2}[t][t]{$r_2$}
\psfrag*{r3}[t][t]{$r_3$}
\psfrag*{1}[r][r]{$1$}
\psfrag*{v}[l][l]{$\vdots$}
\psfrag*{12}[l][l]{$\frac{1}{2}$}
\psfrag*{14}[l][l]{$\frac{1}{4}$}
\psfrag*{18}[l][l]{$\frac{1}{8}$}
\psfrag*{etc}[rb][rb]{etc.}
\psfrag*{a+2}[t][t]{$a+2$}
\psfrag*{a1}[t][t]{$a_1$}
\psfrag*{I1}[b][b]{$I_1$}
\psfrag*{I2}[b][b]{$I_2$}
\includegraphics[keepaspectratio, width=0.9\textwidth]{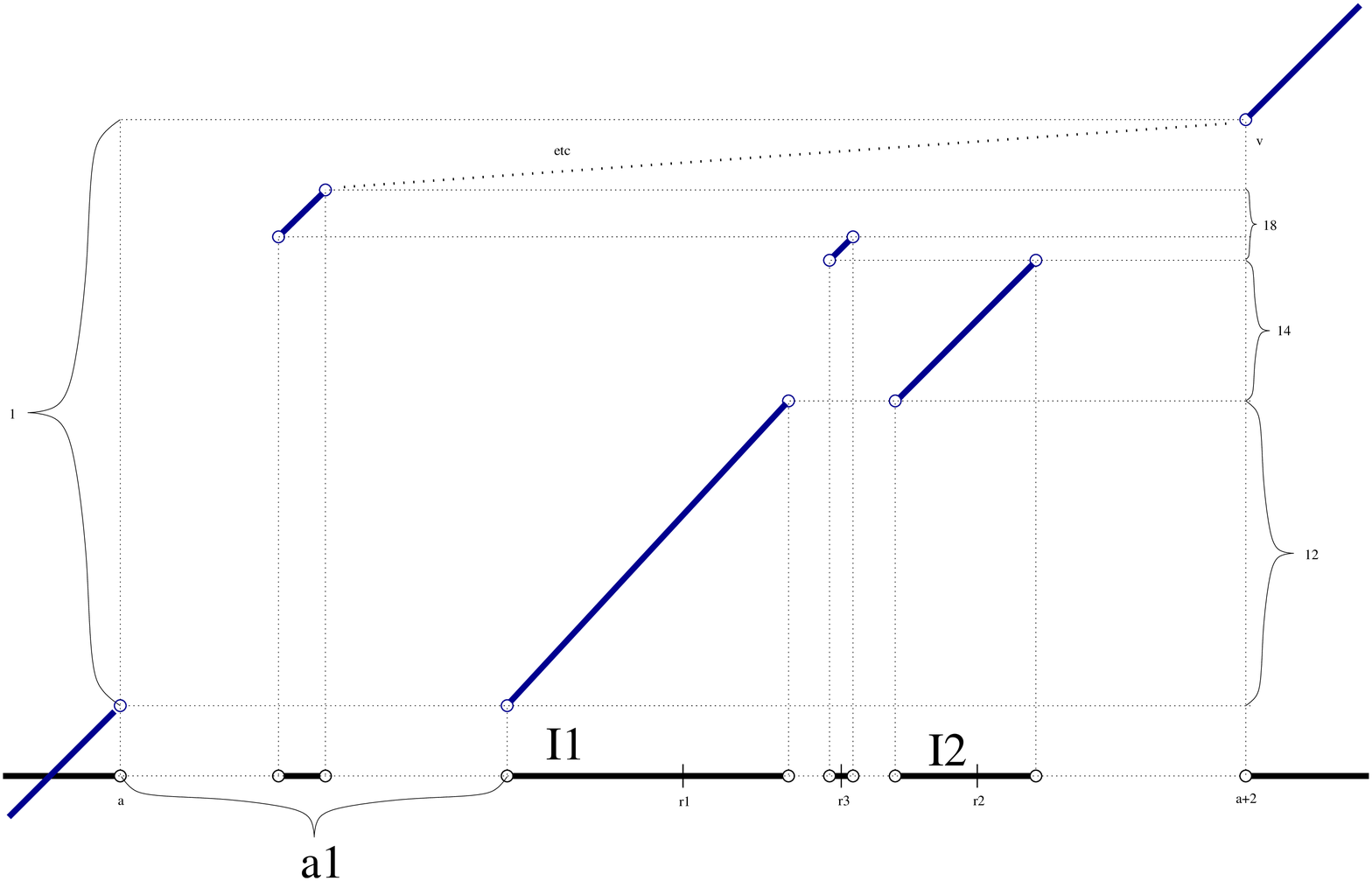}
\caption{\label{arch} for the proofs of Theorems~\ref{thmNoIND} and \ref{thmMO}.}
\end{center}
\end{figure}

To construct the third model, let $\mathfrak{F}$ be an arbitrary countable Archimedean Euclidean ordered field
and let $k(p)=\langle f(p_t),p_s\rangle$ for every $p\in \F^d$
where $f:\F\rightarrow \F$ is constructed as follows,
cf.\ Figures~5, 6.
We can assume that $\mathfrak{F}$ is a subfield of $\mathfrak{R}$ by \cite[Theorem 1 in \S VIII]{Fuchs}.
Let $a$ be a real number that is not an element of $\F$.
Let us enumerate the elements of $[a,a+2]\cap \F$ and denote the $i$-th element with $r_i$.
First we cover $[a,a+2]\cap \F$ with infinitely many disjoint subintervals of $[a,a+2]$ such that the sum of their length is $1$, the length of each interval is in
$\F$ and the distance of the left endpoint of each interval from $a$ is also in
$\F$.
We are going to construct this covering by recursion.
In the $i$-th step, we will use only finitely many  new intervals such that the sum of their length is $1/{2^i}$.
In the first step, we cover $r_1$ with an interval of length $1/2$.
Suppose that we have covered $r_i$ for each $i<n$.
Since we have used only finitely many intervals yet, we can cover $r_n$ with
an  interval that is not longer than $1/{2^n}$.
Since $\sum_{i=1}^{n}1/2^i<1$,
it is easy to see that we can choose finitely many other subintervals of $[a,a+2]$  to
be added to this interval such that the sum of their length is $1/{2^n}$.
We are given the covering of $[a,a+2]$.
Let us enumerate these intervals.
Let $I_i$ be the $i$-th interval, $d_i$ be the length of $I_i$,
$d_0:=0$ and $a_i\geq 0$ the distance of $a$ and the left endpoint of $I_i$.
$\sum_{i=1}^{\infty}d_i=1$ since $\sum_{i=1}^{\infty}{1}/{2^i}=1$. Let
\begin{equation}
f(x):=\left\{
\begin{array}{ll}
x & \text{ if } x<a ,  \\
x-1 & \text{ if } a+2\le x,\\
x-a_n+\sum\limits_{i=0}^{n-1}d_i & \text{ if } x\in I_n
\end{array}
\right.
\end{equation}
for all $x\in \F$, cf.\ Figure~6.
It is easy to see that $k$ is a translation-like bijection.
Let $p,q\in \F^d$ be such that $p_t<a$ and $a+2<q_t$; and let $p':=k(p)=p$,
$q':=k(q)=q-\widehat{1}$ and $m:=m_{o}$.
It is also easy to check that \x{Ddpe} is false in $\mathfrak{M}_{(\mathfrak{F},k)}$ for $k$, $m$, $p$, $q$, $p'$ and
$q'$, cf.\ Figure~5.
\end{proof}

\bigskip
\begin{proof}[\textcolor{proofcolor}{proof of Corollary~\ref{corNoIND}}]
Let  $\mathfrak{F}$  be a field elementarily equivalent to 
$\mathfrak{R}$,
i.e.\ such that all FOL-formulas valid in $\mathfrak{R}$ 
are valid in $\mathfrak{F}$, too. Assume that
$\mathfrak{F}$  is not isomorphic to $\mathfrak{R}$. 
E.g.\  the field of the real algebraic numbers is  
such. Let  $\mathfrak{M}$  be a   model of  
$\x{AccRel_0}$  with field-reduct $\mathfrak{F}$ in which neither  
$\x{Tp}$  nor  $\x{Ddpe}$  is  true. Such  an $\mathfrak{M}$  
exists by Theorem~\ref{thmNoIND}. Since  $\mathfrak{M}\models 
Th(\mathfrak{R})$ by assumption, this
shows  $Th(\mathfrak{R})\cup \x{AccRel_0}\not\models \x{Tp}\lor \x{Ddpe}$. 
\end{proof}

\bigskip

In a subsequent paper, we will discuss how the present methods
and in particular \x{AccRel} and \x{IND} can be used for introducing
gravity via Einstein's equivalence principle and for proving that
gravity ``causes time run slow'' (known as gravitational time dilation). In
this connection we would like to point out that it is explained in
Misner et al.\ \cite[pp.\ 172-173, 327-332]{MTW} that the theory of 
accelerated observers (in flat space-time!) is a rather useful first step 
in building up general relativity by using the methods of that book.

\section*{APPENDIX}

\noindent
A FOL-formula expressing \x{AxSelf^-} is:
\begin{equation}
\begin{split}
\forall m\;\forall p\quad\Ob(m)\wedge \F(p_1)\wedge\ldots\wedge \F(p_d)\;
\Longrightarrow&\\
\Big[\W(m,m,p)\iff \big(\exists b\ \B(b)\wedge\W(m,b,p)\wedge& p_2=0\wedge\ldots\wedge
p_d=0\big)\Big].
\end{split}
\end{equation}

\bigskip
\noindent
A FOL-formula expressing \x{AxPh} is:
\begin{equation}
\begin{split}
\forall m\; \forall p\; \forall q \quad
\IOb(m)\wedge\F(p_1)\wedge \F(q_1)\wedge\ldots&\wedge \F(p_d)\wedge \F(q_d)
\; \Longrightarrow\\
\Big[ (p_1-q_1)^2=(p_2-q_2)^2+\ldots+(p_d-q_d)&^2
\iff\\
 \exists ph\ \Ph(ph)\wedge \W(m,p&h,p)\wedge\W(m,ph,q) 
\Big] \wedge\\
\Big[\forall ph\;\forall\lambda\quad
\Ph(ph)\wedge\F(\lambda)\wedge\W(m,ph,p)\wedge&\W(m,ph,q)\\
\Longrightarrow\;
\W\big(&m,ph,q+\lambda(p-q)\big)\Big].
\end{split}
\end{equation}

\bigskip
\noindent
A FOL-formula expressing \x{AxEv} is:
\begin{equation}
\begin{split}
\forall m\; \forall k\;\forall p\quad
\IOb(m)\wedge\IOb(k)\wedge\F(p_1)\wedge\ldots\wedge\F(p_d)\;
&\Longrightarrow
\exists q\\
\F(q_1)\wedge\ldots\wedge \F(q_d)\wedge 
\big(\forall b\quad \B(b)\; \Longrightarrow\; [\W(m,b,p)\iff&\W(k,b,q)]\big).
\end{split}
\end{equation}

\section*{ACKNOWLEDGEMENTS}

We are grateful to Victor Pambuccian for careful reading the paper and for 
various useful suggestions.  We are also grateful to Hajnal Andr\'eka,
Ram\'on Horv\'ath and Bertalan P\'ecsi for enjoyable discussions. 
Special thanks are due to Hajnal 
Andr\'eka for extensive help and support in writing the paper, 
encouragement and suggestions.

Research supported by the
        Hungarian National Foundation for scientific research grants
        T43242, T35192, as well as by COST grant No.\ 274.

\bibliographystyle{plain}

\end{document}